\documentclass[]{aastex631}

\usepackage{color}
\usepackage{gensymb}
\usepackage{appendix}
\usepackage{hyperref}

\begin{document}

\title{Ion-rich acceleration during an eruptive flux rope event in a multiple null-point configuration}
\author[0000-0003-1790-8018]{Melissa Pesce-Rollins}
\affiliation{Istituto Nazionale di Fisica Nucleare, 
Sezione di Pisa 
I-56127 Pisa, Italy}

\author[0000-0002-3558-4806]{Alexander MacKinnon}
\affiliation{School of Physics and Astronomy, University of Glasgow}

%\collaboration{20}{(AAS Journals Data Editors)}
\author{Karl-Ludwig Klein}
\affiliation{ Observatoire de Paris, LIRA, Univ. PSL, CNRS, Sorbonne Univ., Univ. Paris Cit\'e, 5 place Jules Janssen, F-92190 Meudon, France}
\affiliation{Observatoire Radioastronomique de Nançay, Observatoire de Paris, CNRS, PSL, Université d'Orléans, Nançay, France}

\author[0000-0001-5690-2351]{Alexander Russell}
 \affiliation{School of Mathematics and Statistics, University of St Andrews, Mathematical Institute, North Haugh, St Andrews, KY16 9SS, U.K.}

\author[0000-0001-5685-1283]{Hugh Hudson}
\affiliation{School of Physics and Astronomy, University of Glasgow}

\author[0000-0003-1439-3610]{Alexander Warmuth}
\affiliation{Leibniz-Institut f\"ur Astrophysik Potsdam (AIP), An der Sternwarte 16, 14482 Potsdam, Germany}

\author[0000-0001-6238-0721]{Thomas Wiegelmann}
\affiliation{Max Planck-Institute for Solar System Research, Justus-von-Liebig-Weg 3, 37077 G\"ottingen}

\author[0000-0002-6376-1144]{Sophie Masson}
\affiliation{Sorbonne Université, Ecole polytechnique, Institut Polytechnique de Paris, Université Paris Saclay, Observatoire de Paris, Université PSL, CNRS, Laboratoire de Physique des Plasmas (LPP), Paris, France}
\affiliation{Observatoire Radioastronomique de Nançay, Observatoire de Paris, CNRS, PSL, Université d'Orléans, Nançay, France}

\author[0000-0002-5694-9069]{Clare Parnell}
 \affiliation{School of Mathematics and Statistics, University of St Andrews, Mathematical Institute, North Haugh, St Andrews, KY16 9SS, U.K.}
 
\author[0000-0001-6119-0221]{Nariaki V. Nitta}
\affiliation{Lockheed Martin Solar and Astrophysics Laboratory, Palo Alto, CA 94304, USA}

\author[0000-0002-5448-7577]{Nicola Omodei}
\affiliation{W. W. Hansen Experimental Physics Laboratory, Kavli Institute for Particle Astrophysics and Cosmology, Department of Physics and SLAC National Accelerator Laboratory, Stanford University, Stanford, CA 94305, USA}
 % at an approximate distance of 60~Mm.
\begin{abstract}
We report on the $\gamma$-ray emission above 100~MeV from the GOES M3.3 flare SOL2012-06-03. The hard X-ray (HXR) and microwave emissions have typical time profiles with a fast rise to a well-defined peak followed by a slower decay. The $>$100~MeV emission during the prompt phase displayed a double-peaked temporal structure with the first peak following the HXR and microwaves, and the second one, about three times stronger, occurring $17 \pm 2$ seconds later. The time profiles seem to indicate two separate acceleration mechanisms at work, where the second $\gamma$-ray peak reveals a potentially pure or at least largely dominant ion acceleration. The Atmospheric Imaging Assembly imaging shows a bright elliptical ribbon and a transient brightening in the north-western (NW) region.  Nonlinear force-free extrapolations at the time of the impulsive peaks show closed field lines connecting the NW region to the south-eastern part of the ribbon and the magnetic topology revealed clusters of nulls. These observations suggest a spine-and-fan geometry, and based on these observations we interpret the second $\gamma$-ray peak as being due to the predominant acceleration of ions in a region with multiple null points. The $>$100 MeV emission from this flare also exhibits a delayed phase with an exponential decay of roughly 350 seconds. We find that the delayed emission is consistent with ions being trapped in a closed flux tube with gradual escape via their loss cone to the chromosphere.

\end{abstract}

%\keywords{Classical Novae (251) --- Ultraviolet astronomy(1736) --- History of astronomy(1868) --- Interdisciplinary astronomy(804)}

\section{Introduction} \label{sec:intro}
The signatures of particle acceleration during solar flares are observable over a wide range of energies, with both electrons and ions reaching relativistic energies in some cases. 
The main electron acceleration in a flare is at 10's of keV energies that absorb a significant fraction of the total flare energy \citep[e.g.][]{1976SoPh...50..153L,Ems:al-05}. When accelerated ions reach an energy threshold of $\sim$280~MeV, they can produce neutral pions via p-p reactions, and these decay directly into two $\gamma$-rays. 
This yields a broad continuum peaking at about 70~MeV, 
which may extend up to GeV energies (see~\citeauthor{murp87} \citeyear{murp87} for a review of $\gamma$-ray production in solar flares). Detections of $\gamma$-ray emission from solar flares are not as abundant as those in other wavelengths but there are indications that flare-accelerated ions may carry comparable energy to that of electrons \citep[e.g.\ ][]{Ramaty2000,Emslie:2012aa,Aschwanden2017}. However, due to the limited number of $\gamma$-ray flare detections, one of the unresolved questions regarding ion acceleration during solar flares is whether it shares a close physical relationship with electron acceleration. 

\cite{1988SoPh..118...95V} and \cite{1993AIPC..280..619M}  found correlations between $>$0.3 MeV electron bremsstrahlung continuum emission with 4–8 MeV nuclear excess emission and also with the 2.223 MeV neutron-capture line, while \cite{Chr-90} reported a similar correlation between the $\gamma$-ray fluence in the 4-7 MeV range and the microwave fluence at 15~GHz, produced by the gyrosynchrotron emission of electrons at hundreds of keV. More recently, \cite{Shih_2009} reported a correlation over a range of $>$3 orders of magnitude between the neutron-capture line fluence versus the $>$0.3 MeV bremsstrahlung fluence of solar flares observed with RHESSI \citep{Lin2002} and the $\gamma$--Ray Spectrometer on the Solar Maximum Mission (SMM). 
\cite{Ackermann_2012} studied the impulsive flare SOL2010-06-12 and found that $>$30 MeV $\gamma$-ray emission lags the bremsstrahlung by 6$\pm$3 seconds. All of these findings provide support for a common acceleration mechanism driving the two populations of particles during the prompt phase of the $\gamma$-ray flare emission. 
However, there have also been observations where there was evidence for closely connected acceleration between the 4-7 MeV range of photon energies and hard X-ray emission but at the same time episodes with different time profiles were shown to exist in the same events, as well as $\gamma$-ray peak delays by several seconds \citep{Frr:Chu-83}.

The spatially integrated time evolution of accelerated
electrons and ions during the course of the October 28, 2003 X17 flare showed that the emission from accelerated electrons dominated the initial phase while the $\gamma$-ray line emission from accelerated ions became prominent during the following several hundred seconds~\citep{Oct282003flare}. \cite{Msn:al-09} showed that pion-decay $\gamma$-rays were emitted together with relativistic electrons in the impulsive phase in the January 20, 2005 event, but that the impulsive phase itself consisted of different particle releases with varying ratios of the $\gamma$-ray to hard X-ray or microwave intensities. \cite{1993A&A...275..602C} reported that as the September 9, 1989 flare developed,  the proton acceleration became slightly more efficient than the acceleration of electrons.
Differences between ion and electron acceleration also seemed to be supported by the finding \citep{Hurford_2003,Hurford_2006} that the locations of electron HXR bremsstrahlung and ion 2.2 MeV $\gamma$-ray line emission may differ \citep[see also the review in ][]{Vil:al-11}. But recent work \citep{Battaglia2025} suggests that this result may have been over interpreted.  

The question whether the differences in detailed timing analyses are due to species-dependent propagation and emission, as advocated, e.g., by \cite{Frr:Chu-83} and \cite{Hul:al-92}, or whether they do reveal truly different acceleration scenarios is still open.

The $\gamma$-ray observations of the M3.3 flare of SOL2012-06-03 provide an excellent case study to further investigate the question of whether electron and ion acceleration share a close physical relationship. The time profile of the $>$100 MeV emission detected by the \emph{Fermi} Large Area Telescope~\citep[LAT;][]{LATPaper} during the impulsive phase displays two distinct peaks occurring within the first minute of the flare. The first $\gamma$-ray peak coincides with the 100-300~keV peak whereas the second one does not appear to have any corresponding counterpart in X-rays or radio, suggestive of a potentially pure ion acceleration mechanism. Following the impulsive phase, the $\gamma$-ray time profile exponentially decays for roughly 8 minutes until the Sun leaves the \emph{Fermi}-LAT field of view. In this paper we report on the multiwavelength observations of this flare with the goal to identify the sources of the $\gamma$-ray emission during the prompt and delayed phases.

\section{Gamma-ray observations}

\subsection{Fermi-LAT}
\label{sec:lat_analysis}
The LAT is a pair-conversion telescope sensitive to $\gamma$-rays in the energy range from 30~MeV to $>$300 GeV. It is the main instrument of the \textit{Fermi} gamma-ray space telescope that surveys the entire sky, observing the Sun when it passes through the field of view (roughly 15-20\% of the time).
To analyze the flare SOL2012-06-03 we performed an unbinned likelihood analysis of the \emph{Fermi}-LAT data within the Multi-Mission Maximum Likelihood (\texttt{3ML})\footnote{\url{https://threeml.readthedocs.io/en/stable/index.html}} framework using \texttt{fermitools}\footnote{\url{https://github.com/fermi-lat/Fermitools-conda/wiki}} version 2.0.8. We selected Pass 8 Transient class events from a 10$^{\circ}$ circular region centered on the Sun and within 100$^{\circ}$ from the local zenith (to reduce contamination from the Earth limb). We first analyze the flare integrating over the full duration selecting all the events above 60 MeV to better constrain the shape of the spectrum at low energies and test three models to the \emph{Fermi}-LAT data. The first two, a simple power law (PL) and a power-law with an exponential cut-off (PLEXP)\footnote{The definition of the models used can be found here: \url{https://fermi.gsfc.nasa.gov/ssc/data/analysis/scitools/source_models.html}} are phenomenological functions that may describe bremsstrahlung emission from relativistic electrons. 
The third model  uses templates based on a detailed study of the $\gamma$ rays produced from decay of pions originating from accelerated protons with an isotropic pitch angle distribution in a thick-target model \citep{murp87, Murphy_2009}. In all the three analyses, the background is modeled by a fixed contribution coming from the galactic $\gamma$-ray emission (described by the standard template available in the \texttt{fermitools}), and by an isotropic emission describing the unresolved particle background (also described by the standard available template). This latter background component is left free to vary as it encompasses the background variation due to orbital modulation.
%\hh{\citep{2024arXiv241219586S}}
We rely on the likelihood ratio test and the associated test statistic \citep[TS;][]{Mattox:96} to estimate the significance of the detection. The TS of the power-law fit (TS$_{\rm PL}$) indicates the significance of the source detection under the assumption of a PL spectral shape and the $\Delta$TS=TS$_{\rm ALT}$ - TS$_{\rm PL}$ quantifies how much an alternative model improves the fit. Note that the significance in $\sigma$ can be roughly approximated as $\sqrt{\rm TS}$.

\begin{figure*}[ht!]   
\begin{center}
\includegraphics[width=\textwidth,trim=180 0 180 0,clip]
{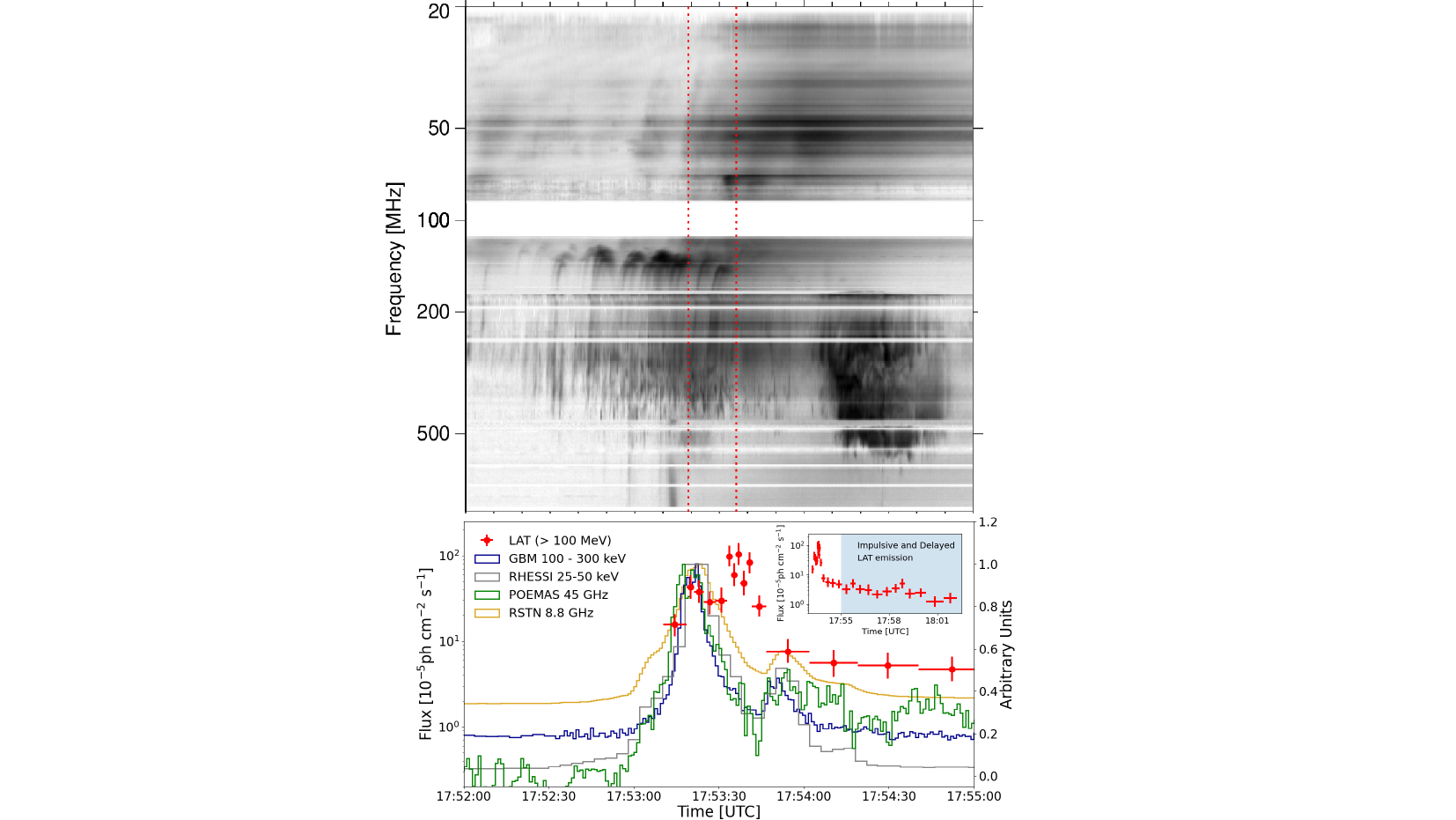}
%{figures/timeprofiles_gbm_lat_rstn.png}
\caption{Multiwavelength light curves of SOL2012-06-03. Top panel: Meter-wave radio combined Bleien-Birr spectrogram. Bottom panel: \emph{Fermi}-LAT $>$100 MeV flux points (red markers, left hand axis) and on the right hand axis are the 100-300 keV GBM time profiles (blue), RHESSI 25-50 keV time profile (gray), radio RSTN at 8.8 GHz in yellow and POEMAS 45 GHz in green (all in arbitrary units). The times in which the $\gamma$--ray flux peaks are 17:53:19 $\pm$1.5~sec and 17:53:36 $\pm$ 1~sec UT (red dashed lines in the upper panel). \emph{Fermi}-LAT observed significant emission from this flare until 18:02 UT (when the Sun left the field of view). The impulsive and delayed \emph{Fermi}-LAT emission, lasting until 18:02 UT, is shown in the insert of the figure. The blue shaded box in the insert indicates the time range of the extended emission of the flare not shown in the bottom panel of the figure.}
\label{fig:timeprofile} 
\end{center}   
\end{figure*}

The impulsive phase of the flare, shown by the red symbols in the bottom panel of Figure~\ref{fig:timeprofile}, displays two peaks, and given the importance of the timing of these peaks, we decided to use the finest time binning possible by requiring that each bin contain at least 12 photons and that the start of the detection (when we have a TS$>$5) coincide with the arrival of the first photon, with a probability greater than 0.9 of being associated with
the Sun. This choice clearly results in limited statistics in each temporal bin and consequently the $\Delta$TS between the PL and PLEXP model is marginal (of the order of 2$\sigma$ in each bin). The time integrated data, on the other hand, are best fit with the curved model with a significance of 6$\sigma$.
The flux values, spectral index and significance of the fit with a power-law model in the time resolved analysis are reported in Table~\ref{tab:spectral_results}. Very little spectral variation is observed throughout the nine minutes of \emph{Fermi}-LAT detection of this flare and no correlation is seen between the spectral index and the flux values.

In order to evaluate the statistical significance of the double peak structure in LAT data we fit the time profile during the impulsive phase of the LAT data with a single gaussian and with a double gaussian. These two simple models are meant to represent the case with a single peak and one with a double peak. The $\chi^2$ and the corresponding p-values are reported in Table~\ref{tab:fits}.

\begin{table}[h!]
    \centering
    \begin{tabular}{c|c|c}
       Model & $\chi^2$/d.o.f & p-value \\
        \hline
         Single gaussian & 37/9 & 3$\times$10$^{-5}$\\
         Double gaussian & 18/7 & 1$\times$10$^{-2}$\\
         \hline
    \end{tabular}
    \caption{ $\chi^2$ over the degrees of freedom and p-values from the fit to the time profile of the LAT data with a single and double gaussian model.}
    \label{tab:fits}
\end{table}

% We then compare this with the value of the mean of the first peak (2012-06-03 17:53:20 $\pm$ 2.5 seconds) found from the fit with the double gaussian to the LAT data. The mean values of these two peaks are consistent within the errors, providing additional support that the electrons and ions producing the emission in the HXR and gamma-rays share a common acceleration mechanism during the time interval covering the first peak. 
rm 
 Based on the results in Table~\ref{tab:fits}, we confirm that there are two distinct peaks during the impulsive phase, the first occurring at 17:53:20 UT $\pm$2.5 sec and a second one at 17:53:36 UT $\pm$ 1 sec. To further test the validity of the first smaller peak in the LAT data, we fit the main peak in the \emph{Fermi}-GBM data between 100 and 300 keV with a single gaussian to obtain the value of the mean, 2012-06-03 17:53:21 $\pm$ 0.5 seconds, which is consistent within the errors with the mean of the first peak in the LAT data suggesting a common origin.

The first peak in the LAT data reaches a flux value above 100 MeV of (43 $\pm$ 14)$\times$10$^{-5}$ ph cm$^{-2}$ s$^{-1}$ whereas the second peak reaches (104 $\pm$ 31) $\times$ 10$^{-5}$ ph cm$^{-2}$ s$^{-1}$. The flux level proceeds to drop by more than a factor of 10 within ten seconds from this second peak. Following this drastic drop, the flux level remains at a $>$5$\sigma$ significance but exponentially decays with an estimated decay timescale, $\tau$ value of 350 seconds until the Sun leaves the field of view of the LAT at 18:02 UT. The \emph{Fermi}-LAT time profile is shown in the bottom panel of Figure~\ref{fig:timeprofile} together with the normalized \emph{Fermi}-Gamma-Ray Burst Monitor~\citep[GBM;][]{Meegan_2009} 100-300 keV light curve, the Reuven Ramaty High Energy Solar Spectroscopic Imager~\citep[RHESSI;][]{2002SoPh..210....3L} 25-50 keV  light curve,  and microwave time profiles observed by whole-Sun radio telescopes at 8.8 GHz \citep[RSTN;][]{Grs:Knw-22} and 45 GHz \citep[POEMAS;][]{Val:al-13}.
 %, RSTN 8.8 GHz and POEMAS 45 GHz time profiles.

%[trim={left bottom right top},clip]

We also performed a time-resolved spectral analysis centered on the individual peaks as well as on the impulsive and the delayed phases of the $>$100 MeV time profile. We define the impulsive phase to take place in the time range 17:53:10 - 17:54:01 and the delayed phase from 17:54:01 - 18:02:11 UT. We found that there is insufficient statistics to be able to distinguish spectral variation between the two peaks in the impulsive phase. In the impulsive and delayed intervals we find that the $\Delta$TS$_{\rm{PLEXP}}$ is greater than 4$\sigma$ indicating that this curved model best describes the data. We also repeated the fit with the pion decay model and found that the preferred proton spectral indices\footnote{ The best proton spectral index is obtained by fitting the data with the pion-decay template models calculated for a range of proton spectral indices.} are $ 5.5 \pm 0.4 $ and $ 4.4 \pm 0.5 $, respectively. With the best proton index and the normalization of the spectral fit, we can get an estimate of the total number of $>$500~MeV protons needed to produce the observed $\gamma$--ray emission over the observed time range, which is $ \left(3.73 \pm 0.17\right) \times 10^{28} $ and $ \left(2.08 \pm 0.23\right) \times 10^{27} $ for the impulsive and delayed phases.

\subsection{Fermi-GBM}
\label{sec:gbm_analysis}
The GBM instrument comprises NaI detectors for hard X-rays and low-energy $\gamma$ rays, plus two BGO detectors; these together cover the range  0.3 - 30 MeV \citep{2009ApJ...702..791M}, i.e. the range in which nuclear line signatures of energetic ions are found. GBM BGO data thus have the potential to give information on flare ion properties complementary to LAT, from the ion energy range $\sim 1 - 100$~MeV \citep{2007ApJS..168..167M}. 

We analysed the data from the GBM B0 detector using the OSPEX software \citep{2020ascl.soft07018T}, initially considering the impulsive period described in Section~\ref{sec:lat_analysis}. We estimated the background by extrapolating from two preflare intervals since the LAT data indicate ion acceleration continuing throughout the extended period up to spacecraft night at 18:02 UT. Although there are some photons above background at $>1$~MeV these are few in number. Spectral fits are poorly constrained but provide interesting upper limits.

\begin{figure*}[ht!]   
\begin{center}
\includegraphics[width=0.8\textwidth]{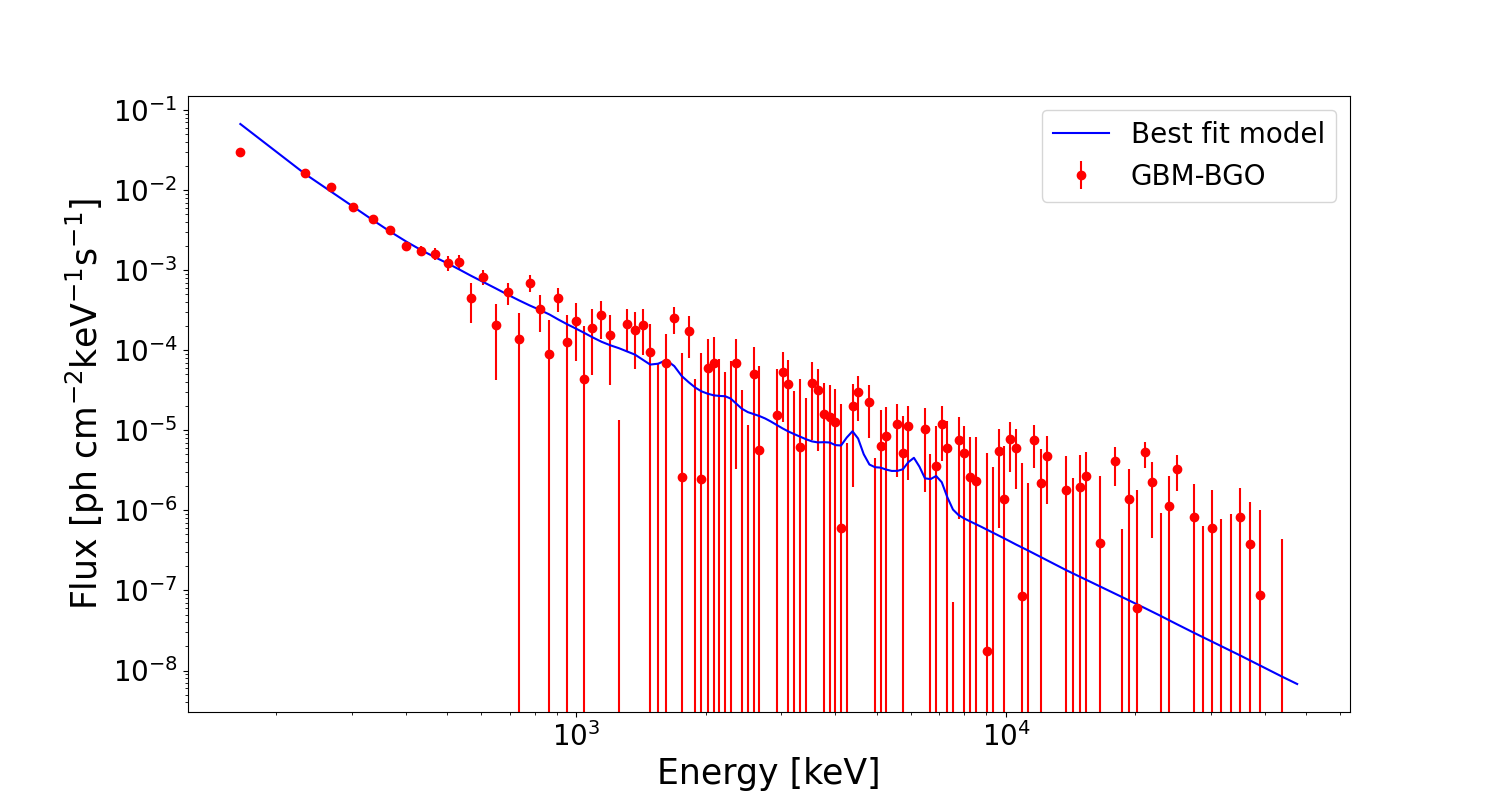}
\caption{Spectrum observed in the GBM B0 (high energy) detector during the interval 17:53:10 - 17:53:46 UT, as a function of photon energy. Also shown is the best fit spectrum combining a broken power--law continuum spectrum and a narrow deexcitation line template, with the parameter values given in the text. }
\label{fig:GBMfit} 
\end{center}   
\end{figure*}

Figure~\ref{fig:GBMfit} shows the GBM B0 spectrum for the interval 17:53:10-17:53:46 UT, together with the best fit to the data for a spectrum comprising three components: broken power-law continuum, narrow line at 2.223 MeV and a narrow deexcitation line template. The neutron capture line is unobservably narrow so we used the OSPEX default line width of 0.1 keV, much less than instrumental resolution. From those available via OSPEX, the narrow line template assumed: a heliocentric angle of $43\degree$; photospheric target abundances and coronal abundances for the fast ions; alpha-to-proton ratio $\alpha/p=0.1$; and a power-law proton energy distribution, $\sim E^{-\delta}$ with $\delta=5.4$. Although the broken power law allows the spectral index $\gamma$ to vary with photon energy, an almost constant value $\gamma = 2.6$ is found throughout the fitted energy range, 0.3 - 7 MeV. Fitting also assigns a negligible amplitude to the 2.223 MeV line and an amplitude of $0.02\pm 0.03$ to the deexcitation line template. An equally acceptable fit ($\chi^2 \simeq 1$ in both cases) is found if we carry out the fit using only power-law continuum. We conclude that there is scant evidence for any nuclear ion signatures at all during the impulsive time interval.

We can treat the estimated amplitude of the line template as an upper limit on nuclear line flux, and thus on proton numbers. Taken at face value it implies a total of $2.1\times 10^{29}$ protons above 30 MeV. If we assume that the $E^{-5.4}$ distribution extends unbroken to 100s of MeV, this implies $8.7\times 10^{23}$ protons above 500 MeV, giving an upper limit orders of magnitude less than the number deduced from LAT (Section~\ref{sec:lat_analysis}). We conclude that the ion distribution interacting with the dense atmosphere must be much flatter or possibly even increasing with energy, below the $>200$~MeV range where pion production occurs. 

We also considered a period immediately following, 17:53:46 - 17:53:56 UT. The neutron thermalization time, several tens of s \citep[e.g.][]{2003ApJ...595L..93M}, leads us to expect 2.223 MeV line appearing at such a later time. The same fitting procedure again yielded an indefinite value for the nuclear line template, but a slightly better defined flux in the 2.223 MeV line: $0.43 \pm 0.27 \mathrm{cm}^{-2}$. With the OSPEX-supplied photon yield for the same assumed ion distribution as above, this fluence implies $(1.7 \pm 1.0)\times 10^{30}$ protons above 30 MeV or $(6.9 \pm 4.3) \times 10^{24}$ protons above 500 MeV - still requiring a flattening of the distribution below the 100s of MeV range.

After this initial period, with its low-significance 2.223 MeV line detection, there is very little signal above background in the BGO detectors. For the whole of the period 17:53:46 - 18:02:00 UT, the same sort of three-component spectrum gives a poor fit to the few counts above background ($\chi^2 = 11$). The amplitude of the 2.223 MeV line is essentially undetermined ($0.006 \pm 0.006$ $\mathrm{s}^{-1}  \mathrm{cm}^{-2}$) while the amplitude of the line template, assuming $\delta=4.4$ as found above from the LAT data, implies $(1.9 \pm 0.6) \times 10^{31}$ protons above 30 MeV. Extrapolating the distribution found from LAT to lower energies implies $(3.0 \pm 0.3) \times 10^{31}$ protons above 30 MeV, so a single power-law distribution from $\sim$~MeV to GeV energies is consistent with both GBM and LAT data during the period following the second $\gamma$--ray peak.

\section{X-ray and radio observations}
\label{sec:xrays_timeprofiles}
The \emph{Fermi}-GBM 100-300 keV time profile is shown in Figure~\ref{fig:timeprofile} together with the \emph{Fermi}-LAT $>$100~MeV fluxes and the microwave time profiles at 8.8 and 45 GHz 
between 17:53 and 17:55 UT and up to 18:02 UT in the insert of the figure. This is the time range over which the flare was significantly detected by the LAT. Two distinct peaks are observed in the hard X-rays, a bright initial peak occurring at 17:53:19 UT and a second minor peak occurring at 17:53:48 UT. The latter is also observed at 8.8 GHz, but not at 45 GHz.  The RHESSI light curves at all energies $\geq$25 keV show the same time profiles. The theory of synchrotron emission, where electrons are assumed to have energy well in excess of their rest energy, predicts that an electron with Lorentz factor $\gamma$ emits up to a frequency of $3/2 \gamma^2 \nu_{\rm ce}$ \citep[e.g., eq. 3.28 of][]{Pac-70}, where  $\nu_{\rm ce}$ is the electron cyclotron frequency. For a magnetic field weaker than 0.4 T emission at 45 GHz implies $\gamma>3$ and therefore a kinetic energy above 1 MeV. This is a lower limit, since the magnetic field and the efficiency of the emission are overestimated. The burst decays to a plateau, which is probably due to thermal bremsstrahlung. The comparison of the light curves shows how the hard X-rays correlate well with the microwave fluxes, as expected from the gyrosynchrotron interpretation of the microwaves. But what is also evident from this figure is how the second $\gamma$-ray peak does not appear to have any evident counterpart in microwaves or X-rays, except for a small bump superimposed on the general decay of the main HXR and 8.8 GHz emission.

\subsection{X-ray imaging}
\label{sec:xrayimaging}
\begin{figure*}[ht!]   
\begin{center}
\includegraphics[width=\textwidth]%{figures/fig_20120603_rhessi_aia_94_wide.eps}
{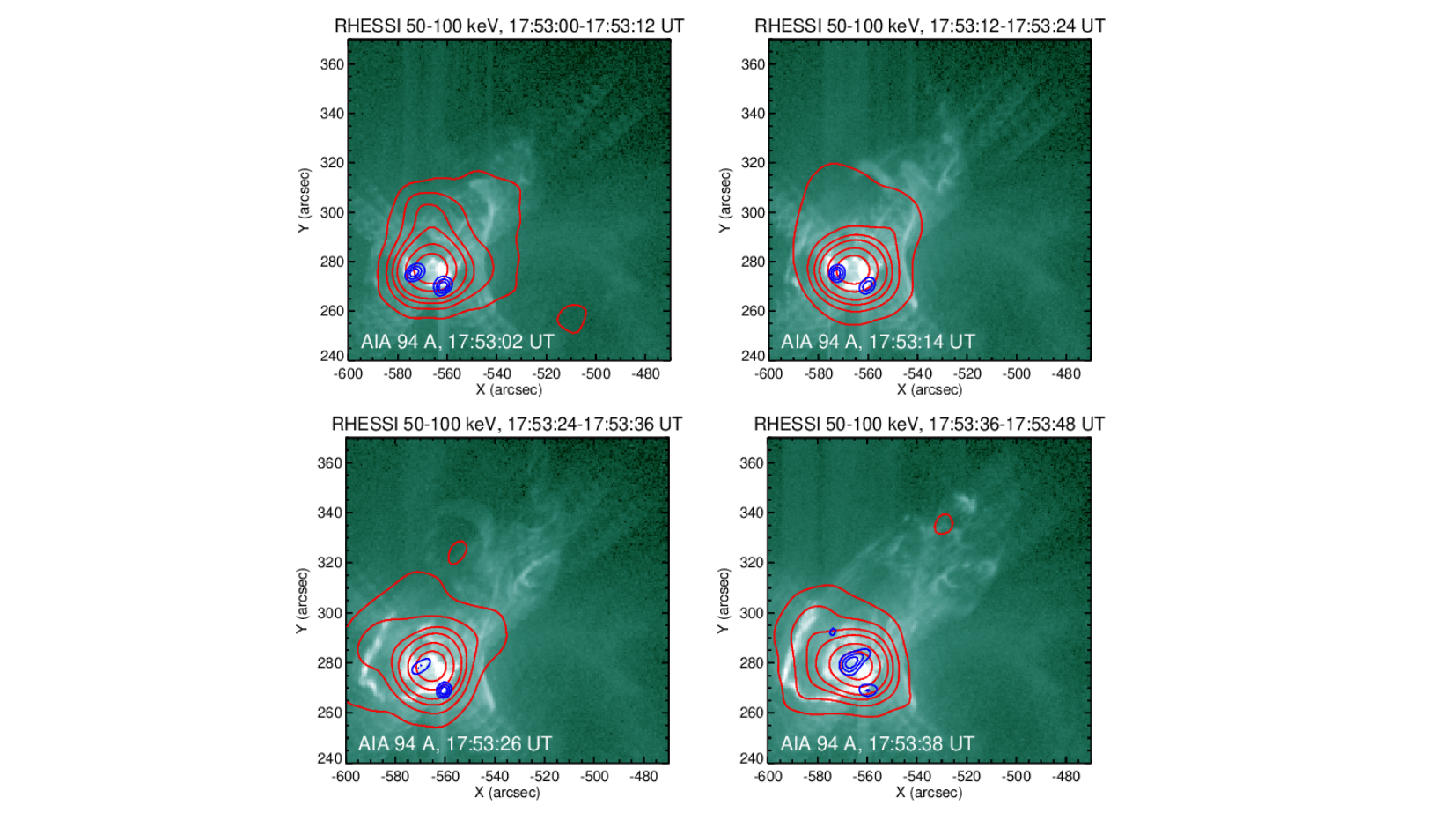}
\caption{RHESSI sources at 12-25~keV (red, at contour levels of 10, 20, 30, 40, 60, and 80\%) and 50-100 keV (blue, contours at  30, 40, 60 and, 80\%) overplotted on AIA 94~\AA\ during the main HXR peak (top two panels) and during the second $\gamma$--ray peak (bottom two panels). Time intervals are reported on the top of each panel.}
\label{fig:rhessi_imaging} 
\end{center}   
\end{figure*}

We reconstructed the nonthermal X-ray sources by applying the expectation-maximization algorithm MEM\_GE \citep{massa2020} to the RHESSI data. Whereas to reconstruct the images in the 12-25~keV range, we used the CLEAN algorithm which gives better results for extended sources as compared to MEM\_GE.
In Figure~\ref{fig:rhessi_imaging} we show the 12 second time integrated 12-25~keV (red contours) and 50-100~keV (blue contours) RHESSI sources overplotted on Atmospheric Imaging Assembly~\citep[AIA;][]{2012SoPh..275...17L} 94~\AA\ images covering the time interval of the main HXR peak and during the second $\gamma$--ray peak. A pair of nonthermal footpoints is clearly visible during the two peaks. During the first peak (top two panels in Figure~\ref{fig:rhessi_imaging}), the footpoint locations show little evolution. However, in the following time intervals (bottom two panels), the northern footpoint becomes stronger and clearly moves to the northwest, indicating that reconnection and acceleration is now active in a magnetic field structure which has a different connectivity than earlier in the event. The thermal emission is concentrated in the southern part of the flare (where also EUV emission is brightest, as will be discussed in Section~\ref{sec:euv}, and the nonthermal footpoints are located). There do not appear to be any other sources with emission above the 10\% level within the AIA region shown in the figure. Note that Fourier-type imagers like RHESSI are limited in their dynamic range to typically less than an order of magnitude.

\subsection{Radio}
\label{sec:radio}

The light curves at hard X-ray and microwave frequencies and the dynamic spectrum in the dm-m-wave range are compared in Figure \ref{fig:timeprofile} with the evolution of the $\gamma$--ray flare. They trace the history of electron acceleration in an extended volume around the flaring active region. The similar time profiles at hard X-rays and microwaves (8.8 and 45 GHz) trace non-thermal electrons with energies between a few tens of keV and at least 1 MeV in the low corona. The light curve at 8.8 GHz evolves smoothly, but it shows discontinuities (near 17:53:00 and 17:53:10 UT, for instance) where the rise accelerates, indicating a succession of acceleration episodes. 

The dynamic spectrum (top panel) shows an enhanced background continuum (grey shading; type IV burst), on top of which spectral fine structure is distinguished: a series of type J bursts between about 450 and 130 MHz and a dense series of bursts with smaller bandwidth between 200 and 500 MHz. The type IV continuum starts near 17:52:30, which is also the time when \emph{Fermi}-GBM sees the first enhancements of the hard X-ray emission. In light curves of the radio emission in the 900-600 MHz range (not shown) the continuum is seen to have a similar time profile as in the microwave range (e.g., 8.8 GHz), but it lasts longer, and the maximum (about 17:54 UT) is increasingly delayed with decreasing frequency. In the dynamic spectrum the continuum is seen down to at least 20 MHz, with a delayed onset at frequencies below 80 MHz. The broad bandwidth of this continuum emission reveals non-thermal electrons in an extended height range, reaching from the bottom of the corona to, say, half a solar radius above. 

On top of the continuum the type J bursts are emitted by electron beams propagating upwards into extended magnetic loops, ceasing emission near the loop tops. The type J bursts start near 17:52:20 UT, 10-20 seconds before the HXR and microwave bursts. They continue through the rise of the HXR and microwave profile. The last J-burst starts near 17:53:30 UT. 

Some of the bursts in the 200-500 MHz range are type III bursts. They are hence also emitted by upward-propagating electron beams. For harmonic radio emission the thermal electron density at the start of the type J bursts (450 MHz) is 6 $\times$ 10$^8$ cm$^{-3}$, the electron density corresponding to the turnover frequency 0.5 $\times$ 10$^8$ cm$^{-3}$. The start frequency of the type J group hence suggests an acceleration region of rather low density, higher in the corona than the regions usually considered as the typical acceleration regions of the hard X-ray emitting electrons \citep[see, e.g., reviews in][]{Asc-02,Kle-21a}, but similar to a simple impulsive flare as reported by \cite{Vil:al-02}. We note that the series of type J bursts ceases 
abruptly during the decay of the main HXR and microwave peak, when the hard X-ray source configuration changes (Sect. \ref{sec:xrayimaging} and Figure \ref{fig:rhessi_imaging}). This commonality corroborates the conclusion that the magnetic connectivity changes at this time, just before the rise to the second $\gamma$--ray peak. Since the type J bursts are emitted in much larger structures than the X-rays, the change of connectivity must be related with the evolution of the coronal magnetic field on large spatial scales. Such an evolution is also illustrated by the EUV images in Figure~\ref{fig:rhessi_imaging}, which show the evolution and eventual disappearance of the erupting flux rope (see Sect. \ref{sec:euv:fluxrope} and Figure~\ref {fig:fluxrope}) on the north-western side of the main EUV brightening and the soft X-ray source. 

The common occurrence of the hard X-rays, type IV-continuum and type J bursts argues in favor of co-ordinated episodes of electron acceleration at different coronal altitudes and the release of non-thermal electrons into a wide range of magnetic structures. While a burst near 17:53:12 UT that drifts from about 500 MHz towards higher frequencies indicates downward-precipitated electrons, the hard X-ray emission and the type IV continuum are well underway at that time. Previous bursts seen in the 500-900 MHz range actually show rapid drifts towards lower frequencies and hence cannot be attributed to downward-propagating electrons. However, these bursts occur at times (e.g., 17:52:57 and 17:53:08 UT) when the microwave continuum emission accelerates its rise. They are therefore rather related to energy release episodes in the low corona, below the region where the electrons emitting the type J bursts are accelerated.

The hard X-ray and 8.8 GHz time profiles have a secondary bump near 17:54 UT. It has no counterpart at dm-m wavelengths and is therefore likely to be due to electrons that remain in the low corona. This event marks also the break of the decay of the $\gamma$--ray light curve, which transits into a slow gradual decay. Higher up, the radio emission evolves now independently of the acceleration processes in the low corona. A new cluster of bursts occurs between 17:54 and 17:55 UT in the range 200-650 MHz, and a type II burst between 18:00 and 18:16 UT below 150 MHz (not shown). The hard X-ray, microwave and $\gamma$--ray emissions show no new enhancement until 18:02 UT, when the Sun leaves the field of view of \emph{Fermi}-LAT 30 minutes.

\section{Extreme Ultraviolet imaging and time evolution}
\label{sec:euv}

\subsection{Overview}\label{sec:euv:overview}
The apparent lack of an X-ray or radio counterpart to the second $\gamma$--ray peak observed during the impulsive phase of this flare suggests an ion-rich particle acceleration. 
To constrain the sources of the high-energy $\gamma$--ray emissions and the locations and mechanisms of particle acceleration, we examined the time evolution of the SOL2012-06-03 flare in all the AIA onboard Solar Dynamics Observatory~\citep[SDO;][]{2012SoPh..275....3P} channels. 

\begin{figure}
    \centering
    \includegraphics[width=0.5\linewidth]{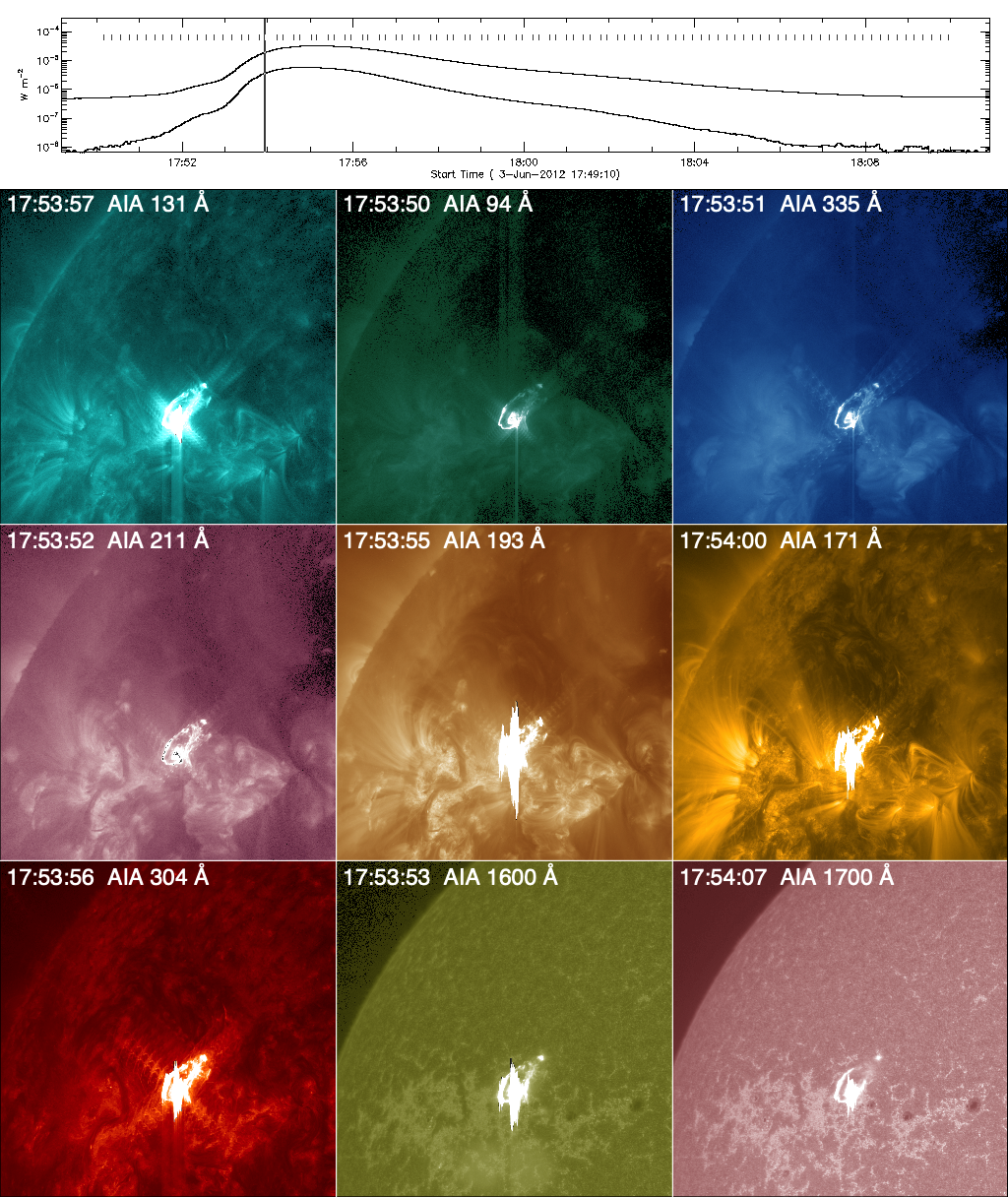}
    \caption{Snapshot of the event in all of the SDO channels. Top panel shows the GOES soft X-ray time profile. The movie of the event is available at  \url{https://www.lmsal.com/nitta/outgoing/fermi_events/20120603/aia_9ch_0304_time_20120603_17.mp4}.}
    \label{fig:snapshot}
\end{figure}

 An overview movie showing all the channels with the maximum 10-s cadence of the instrument is provided as \verb'aia_9ch_0304_time_20120603_17.mp4' and a snapshot of the event is shown in Figure~\ref{fig:snapshot}. 

The light curve at the top of the movie shows the evolution of the whole-Sun X-ray flux in the wavelength ranges 1-8 \AA\ and 0.5-4 \AA\ observed by the GOES satellites. From this movie it is possible see a flare ribbon with an approximately elliptical shape. The eastern part of the ribbon is very bright, while the north-western part has a more complex structure with multiple localized regions of moderate brightness. This is a region of pronounced changes. During the early slow rise of the soft X-rays this is best seen at 131 \AA. During the following steeper rise the brightening continues and becomes visible in a broader region surrounding the compact initial one, and in a wider range of coronal emission lines.  During the final rise to the soft X-ray peak the EUV images show brighter emission with a complex spatial structure: 

\begin{itemize}
\item The flare ribbon evolves into its elliptical shape, mainly clockwise from  south-east to north-west. 
\item As it does, a diffuse broad arc-shaped feature ranging from westward to northward of the flaring active region propagates away from it. Depending on the wavelength, it is seen from 17:53:30 to 17:55:00 UT. We identify this feature as an EUV wave. 
\item On the north-western side of the active region two localized brightenings (henceforth referred to as hot spots) appear near 17:53:30 UT and remain visible at stable positions for about a minute (until 17:54:36 UT at 171 \AA). Both hot spots are faintly visible in the EUV image of Figure \ref{Fig_AIA94}, the northern one is within the yellow rectangle and will be discussed in more detail below. 
\item The two hot spots fade from view when a bright structure connecting them is seen to move north-westwards (e.g., 17:54:45 UT at 131 \AA) and to develop into the geometry of an extended loop, with footpoints that remain rooted in the flaring active region. The feature does not evolve from the hot spots, but appears to come from the bright main flaring region (blue rectangle) and to travel through the region of the hot spots. It is seen in coronal wavelengths and also clearly at 304 \AA. We identify this feature as an erupting magnetic flux rope. 
\item Major structural changes are seen in the northern part of the flaring region around 17:52:23 UT, with the short appearance of a bright loop. They could be at the origin of the outward (north-westward in projection on the disk) expansion of the presumed flux rope.
\end{itemize}

\subsection{EUV intensity around the flare ribbon}\label{sec:euv:intensity}

\begin{figure}[htbp] %  figure placement: here, top, bottom, or page
   \centering
   \includegraphics[width=0.7 \textwidth]{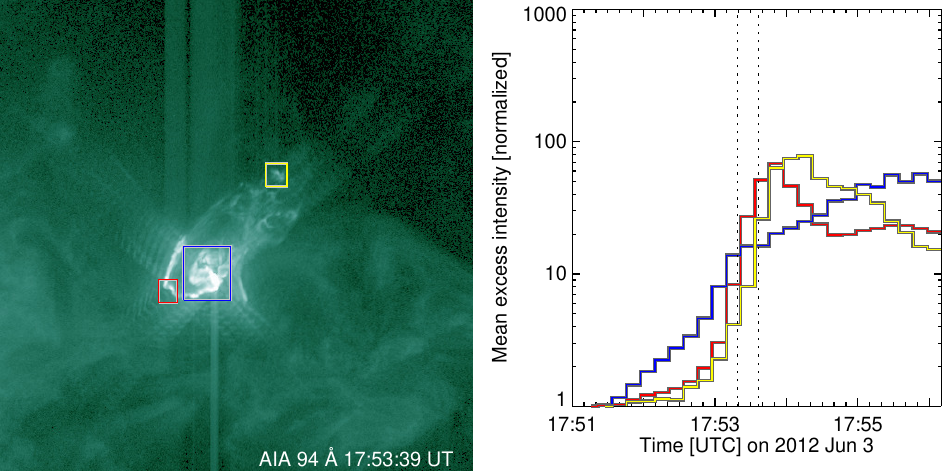} 
   \caption{
Snapshot of the SOL2012-06-03 flare at 94 \AA\ (left panel; the lower left and upper right corners of the image are, respectively, at coordinates [$-718'', 153''$] and [$-394'', 460''$] with respect to the center of the solar disk) and time histories of the average intensity (right panels) in three selected fields delimited by the colored rectangles (the colors of the curves are those of the rectangles). The time histories are normalized to the values at the first instant in the plot. The two vertical dashed lines show the times of the two \emph{Fermi}-LAT $\gamma$--ray peaks.
}  
   \label{Fig_AIA94}
\end{figure}

We analyzed the time evolution of the intensity in selected regions of the flare at 94 \AA\ (Figure~\ref{Fig_AIA94}). At this wavelength, AIA has a significant response to plasma temperatures above $10^{6.5}$ K \citep{Boe:al-12}, and therefore these time histories reveal the evolution of the heating in these areas. We also analyzed images with the 335 \AA\ filter, which has a peak of sensitivity near $10^{6.5}$ K, too, albeit less pronounced and with a higher sensitivity to plasma below $10^6$ K than does the 94~\AA\ filter. 

The movie shows that the northern branch of the ribbon forms gradually between about 17:53 and 17:54 UT, proceeding clockwise from its south-eastern extremity. The light curves in the three rectangular regions overplotted on the AIA image are displayed in the right panel of Figure~\ref{Fig_AIA94} with the same colors as the rectangles. These three regions are (1; blue rectangle)  the most prominent brightening within the flare ribbon, (2; yellow) the north-western hot spot, and (3; red) a region on the south-eastern part of the flare ribbon, to which the magnetic field lines from the north-western hot spot connect, as reported in Sect.~\ref{sec:fieldextrap}. The light curves are the average intensities in the rectangular fields, normalized by their values at the first instant of the plot. The emission of the extended bright region within the elliptical ribbon (blue rectangle and curve) rises gradually until a peak near 17:56 UT. The two other light curves start late, and rise simultaneously in two phases, with an acceleration near 17:53 UT, to maxima near 17:54 UT, i.e. before the peak of the main EUV emission. The maximum in the north-western hot spot (yellow) is slightly delayed, and the decay phases differ. The light curve of the south-eastern part of the ribbon starts a secondary rise at 17:54:50 UT with a peak near the peak time of the main brightening within the ribbon. This statement has to be taken with some reserve since the saturation of the images during the phase of brightest emission may cause a spillover of the signal from the bright region (blue) into the field of the red south-eastern rectangle.

The two $\gamma$--ray peaks, at the times marked by vertical dashed lines, occur during the accelerated rise phase of the emissions from within the north-western and south-eastern rectangles. We examined the rise phase in other rectangular fields along the flare ribbon. The direct neighbors of the south-eastern (red) rectangle outlined in Figure~\ref{Fig_AIA94} display a similar time evolution. More remote fields on the northern part are delayed, while fields on the southern part share the smooth rise displayed by the main brightening. The comparison suggests that the regions in the north-western (yellow) and the south-eastern (red) rectangle comprise footpoints of the magnetic fields to which energy is transferred from a common release region in the corona. This finding meets the expectation from the field-line extrapolation described in Section~\ref{sec:fieldextrap}. The projected distance between the centers of the two rectangles is about 76000~km.

\begin{figure*}[ht!]   
\begin{center}
\includegraphics[width=0.8\textwidth]{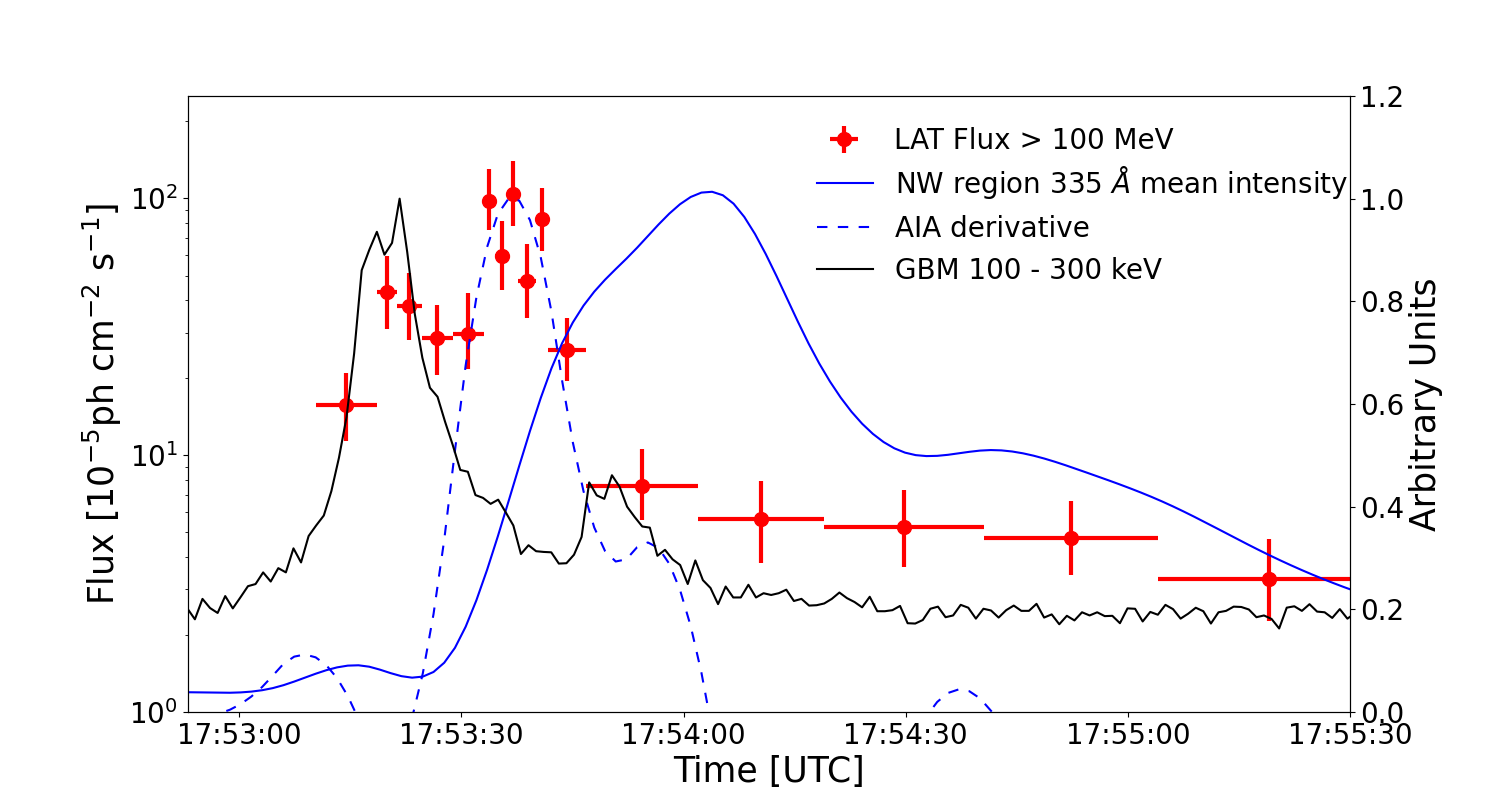}\\
\includegraphics[width=0.8\textwidth]{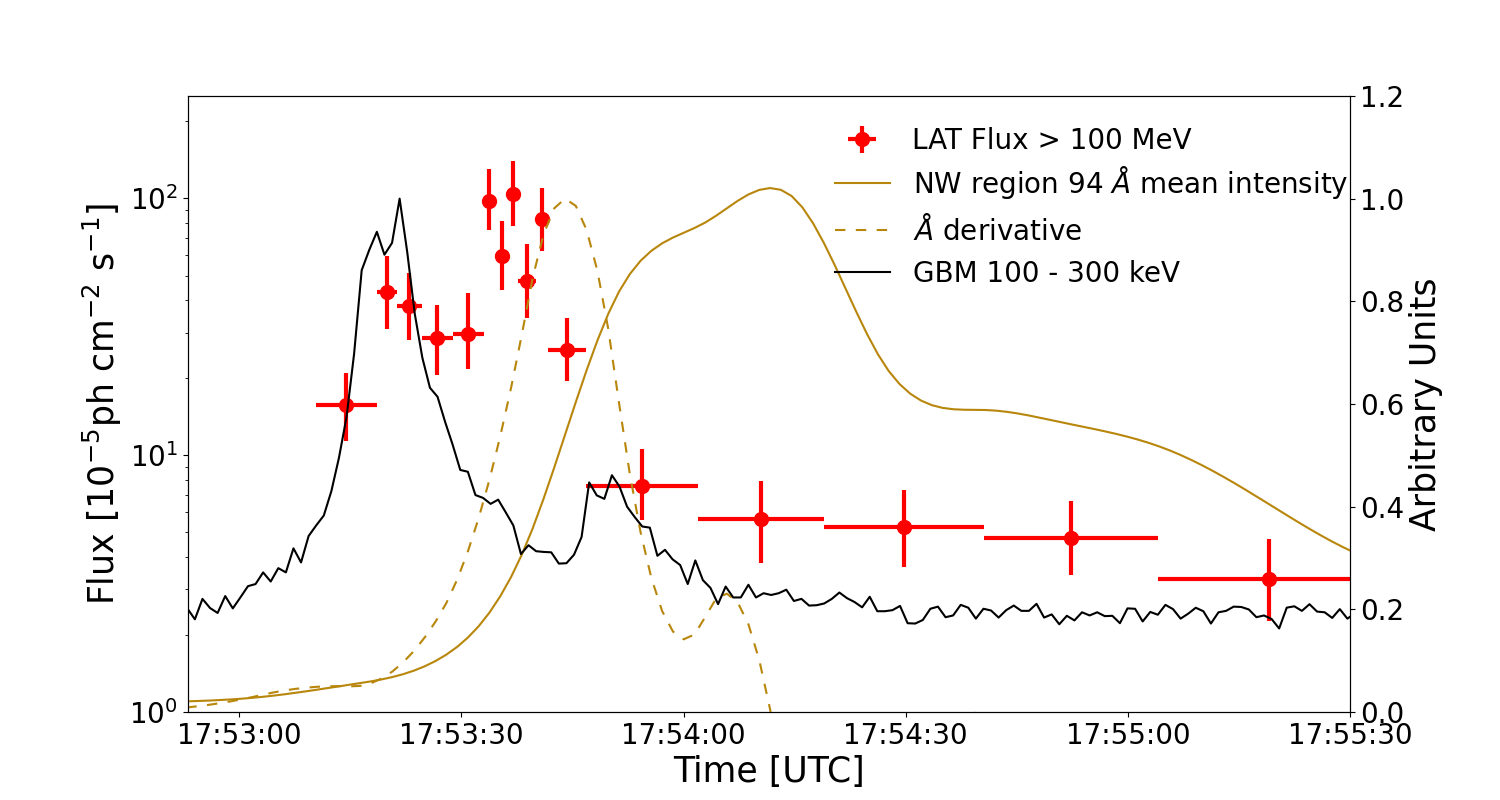}
\caption{Comparison of time histories during the impulsive phase of the flare. \emph{Fermi}-LAT $>$100~MeV flux shown in red, \emph{Fermi}-GBM 100--300~keV counting rates, arbitrarily normalized, shown in black, mean intensity of the north-western region (yellow box in Figure~\ref{Fig_AIA94}) in 335 \AA / 94 \AA\ (top panel/bottom panel) and its time derivative (dashed line). The 335~\AA\ time derivative peaks at 17:53:36 and 94 \AA\ peaks at 17:53:43 UT. The second $\gamma$-ray peak reaches its maximum value at 17:53:36 UT, in coincidence with the 335 \AA\ time derivative and 7 seconds prior to the 94 \AA\, suggesting that the EUV emission observed in the north-western region is physically related to the ion acceleration. The differences in peak time between the 335 \AA\ and 94 \AA\ time derivatives is either due to the variation in diffraction patterns between the two channels or the difference in ions and temperatures that each channel is sensitive to.}
%The times in which the$\gamma$--ray flux peaks, 17:53:19 and 17:53:36 UT, are indicated by the black dashed vertical lines.}
\label{fig:SOL2012-06-03_imp_hxr_euv} 
\end{center}   
\end{figure*}

To better visualize the timing between the second $\gamma$--ray peak and the EUV brightening in the north west region, we describe the AIA 94~\AA\  and 335~\AA\ intensity profiles using a piecewise polynomial function (spline\footnote{The degree of smoothing of the spline is equal to four.}) and compute its derivative. The results are shown in Figure~\ref{fig:SOL2012-06-03_imp_hxr_euv} together with the \emph{Fermi}-GBM and LAT time profiles. From this figure it is possible to see how the time derivative of the AIA 335~\AA\ intensity profile peaks at the same time as the second $\gamma$--ray peak and that of the 94~\AA\ peaks 7 seconds later. The difference in peak times between the two AIA wavelengths can be attributed to the different ion populations and temperatures that each channel is most sensitive to. The occurrence of the $\gamma$-ray peak at the time of the fastest rise of the extreme ultraviolet (EUV) emission is reminiscent of the Neupert effect between thermal X-rays and non-thermal microwaves and X-rays, and similar coupling was observed also for the behind-the-limb $\gamma$-ray flares, as reported in~\cite{Pesce-Rollins_2022}. The timing suggests that the two $\gamma$-ray peaks are associated with energy releases in different parts of the flaring active region and potentially due to different particle populations.

\subsection{Erupting flux rope}\label{sec:euv:fluxrope}
\begin{figure}[htbp] %  figure placement: here, top, bottom, or page
   \centering
   \includegraphics[width= 0.9\textwidth]{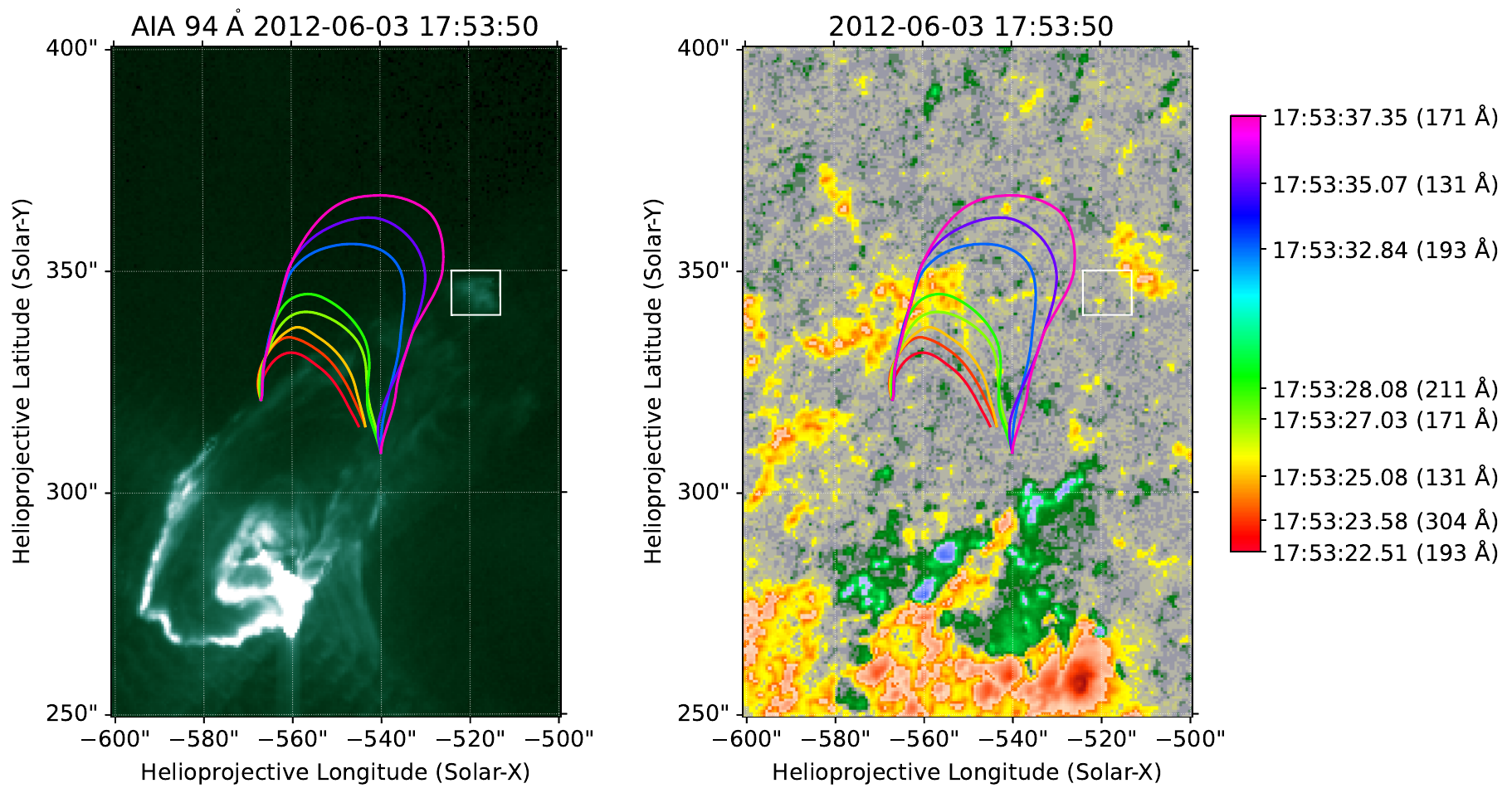}
   \caption{Temporal evolution of the erupting flux rope as traced by the AIA coronal EUV filters overplotted on an AIA 94~{\AA} image and the HMI LOS magnetogram. The color bar reports the midpoint times \texttt{t\_obs} of the AIA exposures used to determine the positions of the erupting loop. The AIA 94~{\AA} image shown in the left panel has \texttt{t\_obs} = 17:53:50 UT.
   %17:53:27.58, hence it shows the flare ribbons at a time when the eruption is underway but the NW hot spot has not yet appeared. 
   %17:53:51.57
   The white boxes indicate the location of the NW hot spot in the AIA 94~{\AA} image.
   %in the later AIA 94~{\AA} image taken at 
}  
   \label{fig:fluxrope}
\end{figure}

An EUV bright loop is visible in all the AIA coronal EUV filters starting from 17:53:05 and remains visible until 17:53:38 UT. According to the dynamics of this event in EUV, this EUV loop is most likely located along a section of the erupting flux rope. Thanks to the fact that the flux rope is visible in multiple filters, we were able to track the eruption at high cadence and find that it evolves towards the NW direction of the flare ribbon. To illustrate the progress of the eruption, we manually identified points along the flux rope in each AIA frame, and interpolated between them using an Akima spline, to overlay the loop at multiple times on one of the AIA images and a line of sight (LOS) of the Helioseismic and Magnetic Imager~\citep[HMI;][]{2012SoPh..275..207S} magnetogram of the region, shown in Figure~\ref{fig:fluxrope}. We note that the interpolated loops are useful to visualize the evolution of the erupting structure,  but do not allow to reproduce the flux rope topology. Loop fits are omitted for 94~{\AA}, 211~{\AA} and 335~{\AA} because the relatively long exposure times for these images produced motion blur and other filters offered sharper images at nearby times. For comparison of timings, the NW hot spot is first detected in AIA images at 17:53:31 UT and the \emph{Fermi}-LAT $>100$~MeV light curve peaks at 17:53:36 UT, i.e. during the later part of the flux rope eruption. The relative timings, locations and directionality strongly support a causal connection in which the flux rope eruption drives the processes that produce the second $\gamma$ ray peak.

\subsection{EUV wave}\label{sec:euv:wave}

\begin{figure}[htbp] 
   \centering
   \includegraphics[width= \textwidth]{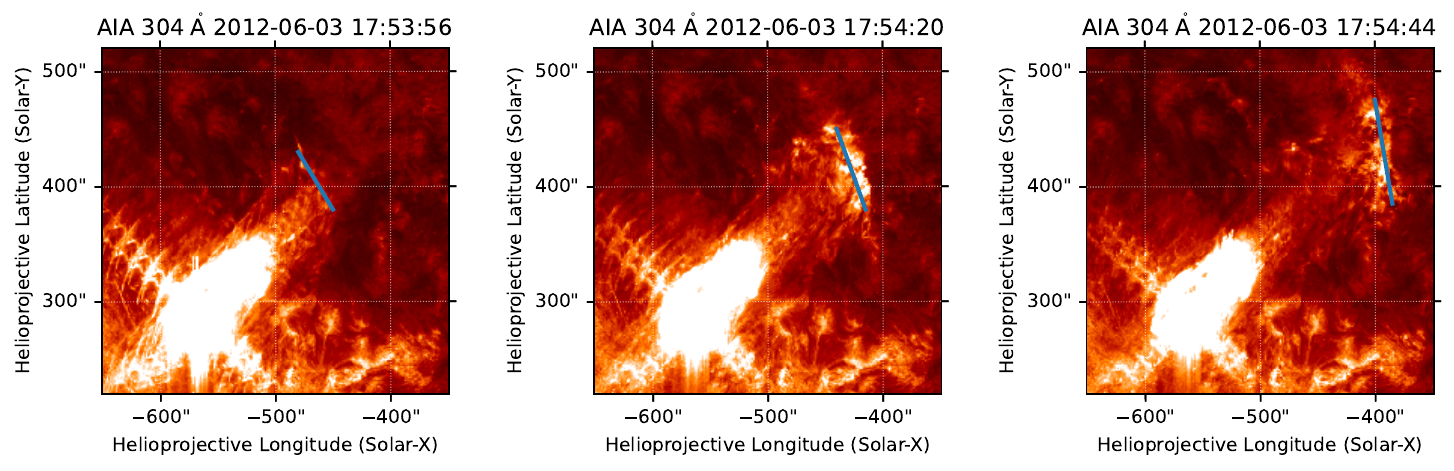}
   \caption{AIA 304~{\AA} wave linked to the second $\gamma$-ray peak. Here, we have shown every second AIA 304~{\AA} image, giving a constant 24~s cadence. The location of the wavefront at each time is indicated by the superimposed straight blue line.
}  
   \label{fig:304_wave}
\end{figure}

Figure~\ref{fig:304_wave} shows that the flare excited a bright wave front in the AIA 304~\AA\ filter channel. It is strongly directed and travels in a NW direction, away from the NW end of the flare. The plane-of-sky speed is $\approx$2000 km/s, implying a high Mach number \citep[cf.][]{Warmuth2015}. The AIA 304~\AA\ channel is sensitive to emission from the chromosphere and transition region, with a characteristic temperature of $10^{4.7}$~K \citep{2012SoPh..275...17L}, therefore we interpret Figure~\ref{fig:304_wave} as showing a wave front at chromospheric altitudes, similar to classic H$\alpha$ observations of Moreton waves. The wavefront is first detected at 17:53:56~UT, in the second 304~\AA\ image obtained after the second $\gamma$-ray peak. Although we do not see the bright front in the immediately preceding 304~\AA\ image, this may simply be because it is obscured by diffraction patterns coupled with the brightness of the flare. Extrapolating backwards in time by 20~s, the propagation is consistent with having originated from the NW hot spot at the time of the second $\gamma$--ray peak.

Propagating disturbances are also evident in the AIA 171~\AA\ and 193~\AA\ channels, with these being most clearly seen in running difference movies. The emission in these channels includes a counterpart to the chromospheric disturbance that travels to the NW. Additionally, coronal intensity perturbations travel large distances towards the east and north-east at speeds of $\approx$800~km/s. The expanding front appears to cause dynamic brightenings at low altitudes near the east limb and north pole (171~\AA\, 17:55--18:05). We also remark that an intensity perturbation in the 193~\AA\ channel leaves the instrument's eastern field of view during 18:02--18:04, propagating away from the Sun.

\section{Nonlinear force-free extrapolations}
\label{sec:fieldextrap}

\subsection{Extrapolation Method}
We used SHARP vector magnetograms from
SDO/HMI \citep{2014SoPh..289.3549B}
and performed nonlinear force-free 
extrapolations that solve
\begin{eqnarray}
(\nabla \times \mathbf{B}) \times\mathbf{B} & = & \mathbf{0}, \label{eq:force-free} \\
\nabla\cdot\mathbf{B} & = & 0, \label{eq:solenoidal}
%
% \label{solenoidal}
\end{eqnarray}
where $\mathbf{B}$ is the magnetic field vector.
These equations have to be solved  with the boundary condition
\begin{equation}
\bf{B}   =  {\bf B}_{\rm obs} \; {\rm on \, the \, bottom \, boundary}.%\nonumber
 \label{Bobs}
\end{equation}

A general overview of force-free solar magnetic fields and
various methods to compute them can be found in the review article by  \cite{2021LRSP...18....1W}.
In the present paper, the input magnetograms were preprocessed to make them force-free consistent as outlined in \cite{2006SoPh..233..215W} and the force-free equations were solved using an optimization code that has been developed specifically for SDO/HMI vector magnetograms, which is described in \cite{2012SoPh..281...37W}. Briefly, to solve the force-free problem, we define a functional
\begin{eqnarray}
L = \int_{V} w_f \frac{|( \nabla \times {\bf B}) \times {\bf B}|^2}{B^2} + w_d \,|\nabla \cdot
{\bf B}|^2 \, d^3V
%\nonumber\\
+ \nu \int_{S} ({\bf B} - {\bf B}_{obs})\cdot{\bf W}\cdot({\bf B} - {\bf B}_{obs}) d^2S
\label{defL}
\end{eqnarray}
The first two terms in the volume integral are zero if the force-free equations 
(\ref{eq:force-free}) and (\ref{eq:solenoidal})
are fulfilled. The $w_f$ and $w_d$ are weighting functions.
The surface integral takes care of the boundary condition (\ref{Bobs}), and the ${\bf W}$ in this term is a diagonal matrix whose elements are inversely proportional to the estimated  measurement errors of the photospheric field vector. The parameter $\nu$ controls the injection speed of the boundary data into the computational domain and $\nu =0.001$ is (usually and also in this case) the optimal choice for HMI vector magnetograms. Further details of the method and code can be found in \cite{2012SoPh..281...37W}. The result of the optimization is the 3D magnetic field vector in the computational domain above the observed active region.

\subsection{Magnetic topology}
\label{sec:topology}

\subsubsection{Analysis methods and visualization}
The 3D structure of the reconstructed magnetic field was analyzed using the Magnetic Skeleton Analysis Tools (MSAT) package, which is described in \citet{benwilliamsthesis} and available from \citet{msat}. This analysis package identifies the 3D magnetic skeleton (i.e. the magnetic null points, their separatrix surfaces and spines, and the separators identified by the intersection of the separatrix surfaces) of the magnetic field by the following steps. First, grid cells containing magnetic null points are identified using the trilinear method \citep{Haynes2007}, which provides definite confirmation that a null point exists within a specific grid cell. The locations of nulls within grid cells are then iteratively refined by applying the trilinear method to sub-grids inside each null-containing cell, until required tolerances have been achieved. The most difficult steps are the classification of the null points (as positive or negative, determined by eigenvalues of the magnetic field in the vicinity of each null) and identifications of the local directions of their spines and separatrix surfaces (determined by the eigenvectors). Full details are given in \citet{benwilliamsthesis}. The separatrix surfaces and spines are then traced outwards from the null points, with care taken to identify where the separatrix surfaces meet null points of opposite polarity. %All opposite-polarity null points that lie on a separatrix surface of another null point are connected to that null point by one or more separators, which are also traced. 
The tracing of the separatrix surfaces and separators are based on the methods used in \cite{Haynes2010}.

The MSAT package also produces 3D visualizations of the magnetic skeleton. In the following figures, magnetic nulls are shown as red and blue balls, and field lines are drawn that include the spines and a sample of field lines in the separatrix surfaces associated to the nulls. Positive nulls, whose separatrix field lines are directed away from the null and whose spine field lines are directed into the null, are colored red, and their fan and spine field lines are colored salmon and red, respectively. Negative nulls, for which separatrix field lines are directed towards the null and spine field lines point away from the null, are colored blue, and their fan and spine field lines are colored pale blue or blue, respectively. Separators, which are special field lines that connect pairs of null points, are shown as thick yellow lines.

\subsection{Full magnetic skeleton}
The MSAT null finder identified 99 magnetic null points in the extrapolated magnetic field of our event. This number is not surprising given the complexity of the photospheric magnetic field distribution, and it is likely typical of solar active regions. The highest null point identified was located 9.1~Mm above the base. However, most of the nulls are low-lying, with 90\% having altitudes below 2.5~Mm, i.e. at heights associated with the chromosphere or photosphere. While the fraction of the null points that are in the corona is therefore relatively small, at 10\%, it is significantly larger than the 2\% fraction reported for the quiet Sun by \citet{2008Regnier} and we attribute the greater null heights in this active region compared to the quiet Sun to the larger scales associated with active region flux distributions. Where null points are found at chromospheric heights, departures from non-force-free (low beta) conditions are likely to change the locations of the null points, however, we believe that the large scale topology of the active region, which our discussion focuses on, is robust.

\begin{figure}
   \centering
   \includegraphics[width=\textwidth]{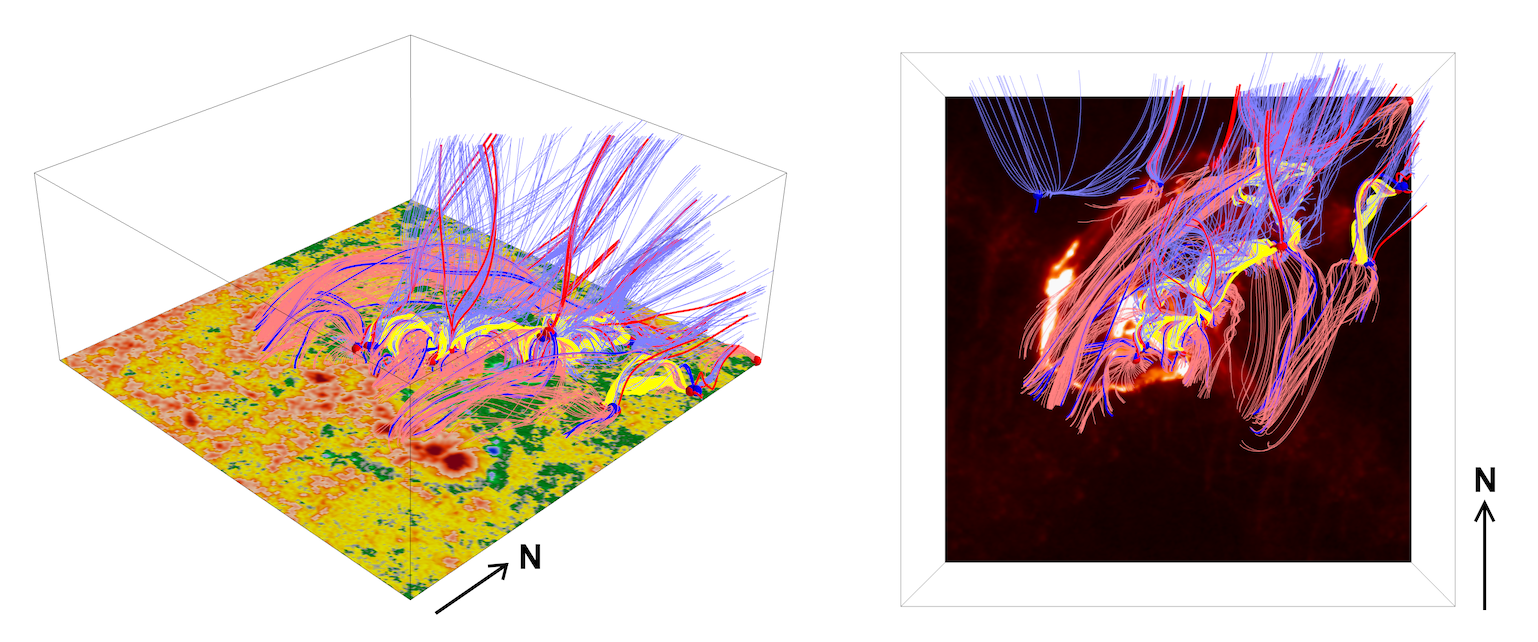} 
   \caption{Magnetic skeleton of the reconstructed magnetic field shortly before the flare and its relation to the flare ribbons. (left) Perspective view showing the SHARP $B_r$ at 17:48:00 on the base, using the standard colorized HMI color table. (right) Top view showing flare ribbons in AIA 304~{\AA}. The skeleton is visualized by showing null points as red or blue balls, spine field lines (thick red and blue lines), fan field lines (thin salmon-colored or pale blue lines) and separators (thick yellow lines); see also Figure~\ref{fig:topology:spot} for greater clarity. Reddish/blueish color identifies positive/negative 3D nulls and their associated field lines. The size of the base is 304 SHARP pixels in both dimensions, making the sides of the base approximately 111 Mm.}
   \label{fig:topology:all}
\end{figure}

Figure~\ref{fig:topology:all} shows the full magnetic skeleton from two viewpoints, with HMI SHARP $B_r$ and the AIA 304~{\AA} emission shown on the base. The AIA data were reprojected onto the SHARP patch coordinates using {\tt sunpy}\footnote{\href{https://sunpy.org/}{https://sunpy.org/}} and the alignment was manually fine-tuned by aligning sunspots in AIA 1700~{\AA} with the corresponding $B_r$ concentrations, which resulted in a translation by a few pixels.
Examining Figure~\ref{fig:topology:all}, a striking topological feature is the large cluster of interconnected nulls that runs along the western edge of the flaring region, most easily seen in the right panel as the network of yellow separators along the western edge of the elliptic ribbon. The flare starts at the southern end of this null cluster network, and it spreads toward the northern end where the NW hot spot is located, as more of the null point network becomes activated. Comparing Figures~\ref{fig:rhessi_imaging} and \ref{fig:topology:all} shows that the main X-ray source and much of the brightest EUV emission remain at the southern end of the null point network.

The magnetic skeleton of the extrapolated magnetic field explains some of the major visual differences between different parts of the flare ribbons. The south and south-eastern parts of the ribbons are narrow and bright, and these coincide with locations where fan surfaces intersect with the base (as is illustrated by the salmon colored lines in Figure~\ref{fig:topology:all}), analogous to the circular ribbon of a simple circular ribbon flare. In contrast, the west and north-western parts of the elliptical flare ribbon are spread over a larger area with a broken appearance. The extrapolation indicates that this appearance reflects the topological complexity in this part of the active region, with the dispersed ribbons being associated with magnetic footpoints of an extensive cluster of nulls  (traced by the thick yellow lines in Figure~\ref{fig:topology:all}). Finally, we remark that when comparing the flare ribbons to the magnetic skeleton, it should be borne in mind that AIA 304~{\AA} has a characteristic temperature of $10^{4.7}$~K \citep{2012SoPh..275...17L}, which is expected to occur several Mm above the photospheric base. This can produce an offset between the intensity image and the corresponding magnetic field lines, with this effect being largest where the field is highly inclined. The advantage of showing AIA 304~{\AA} instead of 1600~{\AA} or 1700~{\AA} (which form closer to the base) is the availability of a short exposure taken during the flare that avoids saturation and bleeding and, therefore, most clearly shows the flare ribbon structure. 

\subsubsection{Topology of NW hot spot region}\label{sec:hot-spot-topology}

\begin{figure}
   \centering
   \includegraphics[width=\textwidth]{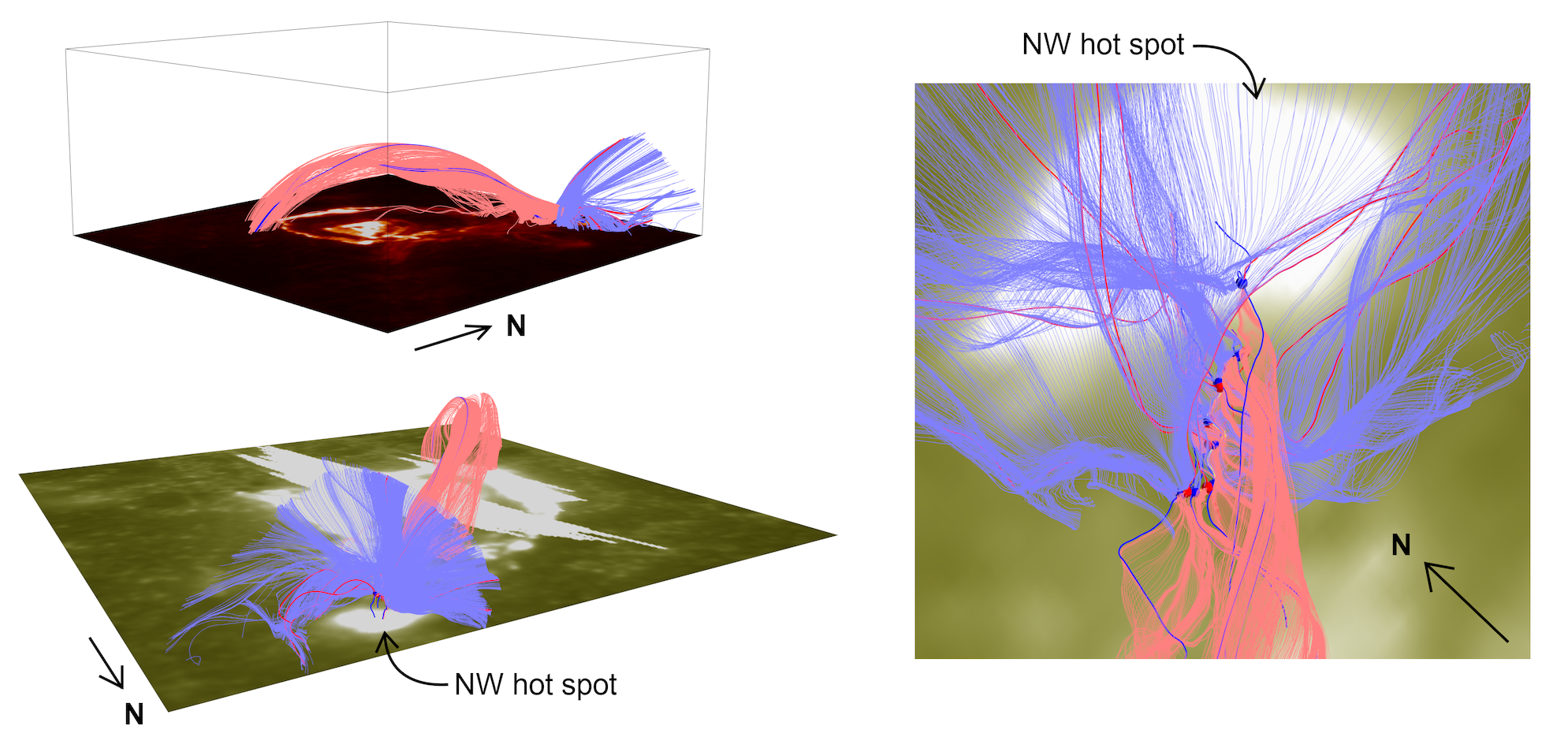} 
   \caption{Magnetic skeleton components most relevant to the NW hot spot, visualized as for Figure~\ref{fig:topology:all}. (upper left) Side view from the SW showing flare ribbons in AIA 304~{\AA} on the base. Blue fan field lines and red spines form a dome-like structure at the NW corner, while salmon-colored fan field lines and blue spines form a wall that arches over the active region, connecting in the SE ribbon. (lower left) View from the NW into the blue dome, showing the hot spot emission in AIA 1600~{\AA}. The hot spot coincides with the footpoints of blue spine field lines from negative nulls. (right) A slightly-inclined top-down close-up of the null point chain, where the blue dome and salmon-colored wall intersect. The spine field line that connects most centrally into the NW hot spot belongs to a negative null point that is located at the NW end of the chain of nulls (represented by the red and blue balls).}
   \label{fig:topology:spot}
\end{figure}

The main focus of this paper is the ion-rich particle acceleration of the second $\gamma$-ray peak, which the observations indicate is somehow connected with the NW hot spot. We therefore focus now on the magnetic skeleton components related to the magnetic nulls that are near to the NW hot spot, shown in Figure~\ref{fig:topology:spot}. With the rest of the skeleton stripped back, Figure~\ref{fig:topology:spot} (upper left) shows that the reconstructed magnetic field contains a dome formed of blue fan and red spine field lines, in the NW corner, threaded by a wall comprised of salmon-colored fan and blue spine field lines that arch over the active region, connecting to the photosphere in the SE part of the flare ribbon.

Figure~\ref{fig:topology:spot} (lower left) shows a view looking into the dome from the NW. During flares, the AIA 1600 and 1700~{\AA} filters are dominated by emission from continuum and spectral lines formed at chromospheric and transition region temperatures \citep{2019ApJ...870..114S}, making these AIA images especially well suited for comparing flare ribbons with magnetic topology. The 1600~{\AA} image taken at 17:53:53 UT captures the NW hot spot well, with the hot spot unaffected by the saturation and bleeding that affects other parts of the ribbons, hence this image is shown on the base in Figure~\ref{fig:topology:spot} (right). The NW hot spot directly coincides with the footpoints of dark blue spine lines that are connected with negative null points.

The topological structure we have isolated in Figure~\ref{fig:topology:spot} is similar to the fan and spine skeleton of a single negative null point, with the hot spot situated at the footpoint of a spine field line. The dome and wall structure can be interpreted by considering what happens if a single negative null point splits into a chain of $n+1$ negative null points and $n$ positive nulls, alternating in sign along the chain. In this case, the original spine field line becomes a wall that is bounded by a pair of negative spines, which corresponds well to the structure we have identified. \citet{benwilliamsthesis} provides an extended discussion of topological structures like these and their occurrence in the solar atmosphere. 

Figure~\ref{fig:topology:spot} (right) provides a closeup of the reconstructed null chain, using a slightly inclined (15\degree) top-down view. The blue spine that connects into the center of the hot spot belongs to the negative null point (blue ball) at the NW end of the null chain. This null point's spine field lines form one edge of the wall of salmon colored fan field lines. It is likely to be pertinent that the chain of null points runs from SE to NW, which aligns with the direction of the flux rope eruption (Figure~\ref{fig:fluxrope}) and the direction in which the AIA 304~{\AA} wave front propagates (Figure~\ref{fig:304_wave}). This alignment may be potentially important for the acceleration. Spines from nulls at the SE end of the null chain connect into a fainter patch of emission, near the bottom edge of the image.
Null points in the chain have low altitudes, consistent with the upper chromosphere or transition region, where the relatively large density means that in principle acceleration benefits from a large supply of particles but is impeded by high collisionality. Finally, we do not put too much weight on the precise number or altitude of the null points, which may be subject to uncertainties from the field extrapolation or may change during the dynamics of the eruption and flare, however, we believe that the existence of an extended weak-field region with a SE-NW orientation is robust, as is the large-scale magnetic topology.

The main conclusions from this topological analysis are that the NW hot spot and its conjugate brightening in the SE ribbon are consistent with particle acceleration occurring in the null cluster at the NW corner of the active region, and that the particles that power the emission of the hot spot precipitate along the spines of the negative nulls (and neighboring field lines). The direction of the null chain suggests that ions may be accelerated by a shock wave passing through this extended weak field region, which then precipitate at the far end of the null chain. Alternatively, the acceleration may be associated with onset of reconnection within null cluster, e.g. in response to stressing of the null points by the flux rope eruption.
%{\rd In Section~\ref{sec:interp_impulsive} we tend to lean more towards the interpretation that the second$\gamma$--ray peak is due to magnetic null reconnection triggered by the erupting flux rope and not due to the shock wave. Should we modify this last sentence to reflect this?}

\section{Interpretation}\label{sec:interpretation}

\subsection{The impulsive phase}
\label{sec:interp_impulsive}

\begin{figure*}[ht!]   
\begin{center}
\includegraphics[width=\textwidth]{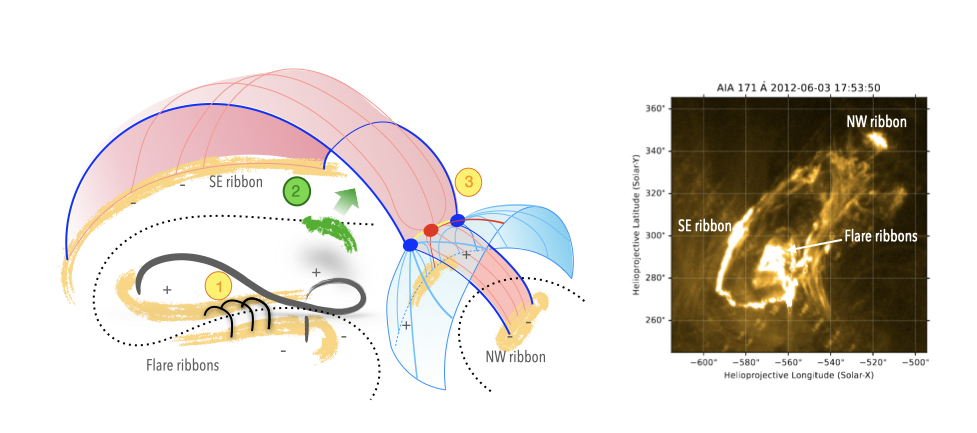}
\caption{Geometry of the magnetic field including two fan surfaces and three null points, one surface delimiting the flare ribbon (large pink dome) and one surrounding the NW ribbon (smaller blue dome).  The labels represent three key features tied to the particle acceleration during this event. Label 1 indicates the region where the particle acceleration leading to the first peak is believed to occur, label 2 shows the EUV bright loops associated with the erupting flux rope driven-shock and label 3 is the location of the cluster of magnetic nulls near the NW hot spot. The right panel shows the AIA 171 \AA\ image of the elliptical ribbon with the three regions indicated.}
\label{fig:cartoon} 
\end{center}   
\end{figure*}

From the magnetic configuration presented in Section~\ref{sec:topology}, one possible scenario that could explain the impulsive phase of this flare is illustrated in Figure~\ref{fig:cartoon}. The particle acceleration leading to the 
first peak appears to occur during the main flare phase in the vicinity of the J-shaped flare ribbons labeled ``1'' in the figure.  
This is supported by the similarities in the time profiles between $\gamma$~rays and emissions in hard X-rays and radio waves (shown in Figure~\ref{fig:timeprofile}), as well as the RHESSI footpoints shown in Figure~\ref{fig:rhessi_imaging}.
The predominant emission of ions in the second peak is a strong argument for an acceleration process at a different place and possibly of a different nature from the one accelerating the electrons and ions in the first peak.  The change of location of the associated energy release is supported by the temporal association between the $\gamma$-ray peak and the distinct EUV brightening on the north-western side of the elliptical flare ribbon. The change of the dynamic spectrum in the 500-100 MHz range, with the time coincidence of a series of type J bursts with the main HXR and microwave peak, together with the first $\gamma$-ray peak, and the absence of these bursts during the second $\gamma$-ray peak, further supports this idea. This is not the first case where accelerated electrons and ions displayed distinct and separate phases during a flare. \cite{Oct282003flare}, for example, reported significant changes in the time evolution of the nuclear line emission with respect to the bremsstrahlung emission during the flare of October 28, 2003. However, this is the first time that such distinct phases are observed for $>$100~MeV emission and that the location of the energy release and the acceleration process for the second $\gamma$-ray peak are identified as being separate from the first one.

%The spine-and-fan scenario, as envisioned by \cite{2009ApJ...700..559M}, suggesting two distinct sites of magnetic reconnection; we speculate that the impulsive $\gamma$-ray bursts come from the null-point region. Here QSL refers to quasi-separatrix layer, and HFT to the ``hyperbolic field line'' giving rise to the double-ribbon geometry.

\subsection{Acceleration mechanisms}
The time evolution of the second $\gamma$-ray peak is impulsive, like the first one, and the coincidence with the rise of the EUV intensity confirms the similarity with impulsive acceleration processes. The particle acceleration in this part of the flare is unusually ion-rich and it coincides with the eruption of a fast magnetic flux rope (grey line anchored in the J-shaped flare ribbons in Figure~\ref{fig:cartoon}). The topological analysis of the magnetic field in Section~\ref{sec:hot-spot-topology} indicates that particles producing the NW hot spot are accelerated in a cluster of magnetic null points that are interconnected by magnetic separators. Here, we consider two main types of explanation: acceleration by reconnection (by DC and Fermi mechanisms, at the reconnection site labeled 3 in Figure~\ref{fig:cartoon}), and acceleration involving a shock driven by the eruption (shown by the green arc in Figure~\ref{fig:cartoon}). It is also possible that mechanisms of both types could work together, with one mechanism providing the initial injection of energetic particles and another process further continuing the acceleration energization.

When the flux rope erupts beneath the ``magnetic wall'' identified in Section~\ref{sec:hot-spot-topology}, the magnetic field in the vicinity of the null cluster will naturally be stressed by the eruption, which we would expect to form intense parallel currents along the separators. This raises the possibility of acceleration by at least two reconnection mechanisms. Firstly, if parallel currents concentrated at the separators have an associated parallel electric field, e.g. due to resistivity, then particles can be accelerated by a DC mechanism. \citet{2016A&A...585A..95T} and \citet{2020A&A...635A..63B} have recently modeled this process using MHD simulations and test particles, obtaining proof-of-principle that this mechanism may accelerate particles into the MeV to GeV range, especially given stronger current density and longer separators. The main objection to DC acceleration is that it cannot explain preferential ion energization. In fact, a hypothesis of acceleration by DC electric fields implies that electrons should also be accelerated to greater than 100~MeV, which we do not see evidence for.

The second reconnection mechanism that we consider is Fermi acceleration. It is well established that thin current layers fragment by tearing instabilities; see for example \citet{2014PhPl...21h2114W} for a simulation of nonlinear tearing of a current sheet at a 3D null point. The nonlinear development following the instability produces conditions favorable to acceleration by Fermi processes. \citet{2009JGRA..114.5111D} and \citet{2014PhPl...21g2903D} have argued that ions are preferentially energized during reconnection because ions gain an immediate kick of order $m_iV_A^2$ when they enter a reconnection exhaust (whereas electrons gain much less energy in this initial step) and subsequent acceleration within merging flux ropes causes particles to gain energy at a rate proportional to the particle energy \citep{2006Natur.443..553D, 2013ApJ...763L...5D}. The ions therefore run further ahead in energy space and the electrons never catch up. This scenario has been validated in 2D simulations \citep[e.g.][]{2024ApJ...974...74Y} and 3D simulations \citep[e.g.][]{2014PhPl...21i2304D,2017PhPl...24i2110D,2021PhRvL.127r5101Z,2022ApJ...936L..27C}.

Alternatively, the second $\gamma$-ray peak coincides with the eruption of a magnetic flux rope. Because of its high speed it may drive a shock wave, especially in the restricted region of weak magnetic field around the null points. First-order Fermi acceleration at shocks is a well established mechanism that has been proposed to accelerate particles during solar flares~\citep[e.g.][]{rank2001,Share_2018, flarecatalog_2021, cliv93,Pesce-Rollins_2022}. The limited duration of the shock could in principle explain the short duration of the $\gamma$-ray peak, the spatially localized NW hot spot in EUV and its stable position, and the correlated EUV-brightenings of the north-western and south-eastern ribbons (yellow and red rectangles in Figure~\ref{Fig_AIA94}), which support instead a scenario of localized energy release on field lines connecting the conjugate locations.

In this scenario the duration of the second $\gamma$-ray peak should be similar to the time it takes for the shock to transit the weak-field region. In the plane-of-the-sky approximation, the length of the null cluster is about 5000~km, and the EUV wave that is later seen propagating away from the NW hot spot travels at around 2000~km/s (Section~\ref{sec:euv:wave}), implying that the rise time should be approximately 2.5~s. This time scale is consistent with the rapid increase in the \emph{Fermi}-LAT $>100$~MeV flux at the onset of the second peak, seen in Figure~\ref{fig:timeprofile}.

%A more suitable hypothesis is that

In this hypothesis particles would  be accelerated by repeated reflections as the shock passes through the weak-field region associated with the NW hot spot, which acts as a magnetic trap. Here, we envisage a shock advancing NW along the chain of null points visualized in Figure~\ref{fig:topology:spot} right. Particles that are trapped within the weak field region bounce back and forth between the magnetic mirror at the NW end of the chain and the shock, equivalent to alternate reflections between a static mirror and a moving mirror. When particles reflect from the shock (the moving mirror) their speed normal to the shock increases by twice the speed of the shock. 

This mechanism is, however,  inefficient because the energy gain is along the direction of the magnetic field and leads to a rapid escape from the trap. The relative energy gain is comparable to the magnetic mirror ratio, and stays well below the 300 MeV required for the $\gamma$-ray emission. Scattering back to the shock therefore needs the development of upstream turbulence, which is advocated in numerical models of diffusive shock acceleration \citep[e.g.,][]{Afa:al-18b}. It is uncertain that such a process can develop on the short time scales imposed by the observations.

\subsection{The delayed phase}
\label{sec:intmed}

The radiation observed by LAT shows a steady decline from about 17:54. 
This time behavior is consistent within errors with an exponential decay, $\sim e^{-t/\tau}$ with decay timescale $\tau \approx 350$~s. A simple interpretation suggests itself. 
Some of the ions accelerated by the events of the flare remain trapped on closed field lines, in a low-density region where they lose energy and radiate negligibly. 
They escape gradually to the chromosphere and photosphere where they stop promptly, emitting $\gamma$~rays in the process. 
In a first approximation the observed decay time of $\gamma$-radiation will be identical with $\tau$. 
We can test the plausibility of this idea.

Let $N(E,t)$ be the distribution in energy (GeV$^{-1}$) of accelerated protons of energy $E$ at time $t$ in this reservoir. Neglecting in the first instance any energy-dependence of $\tau$ we have 
\begin{equation}
    N(E,t) \, = \, N_0(E)E^{-t/\tau}
\label{eqn:net}
\end{equation}
\noindent for some energy distribution $N_0(E)$, to be determined from the $\gamma$--ray spectrum. The flux of protons into the dense atmosphere at time $t$ is given by $N(E,t)/\tau$. 

We find that the LAT data in the period 17:54:01--18:02:11~UT are best fit assuming $ \left(2.08 \pm 0.23\right) \times 10^{27} $ protons above 0.5~GeV, with a power-law energy distribution of energy spectral index $\delta=4.4\pm 0.5$. In this 480 second period, 0.75 of the initial proton population has escaped to emit $\gamma$~rays. 
Thus
\begin{equation}
    N_0(E) \, = \, A \left(\frac{E}{0.5}\right)^{-4.4}
\label{eqn:n0def}\end{equation}
\noindent where $A = 1.81\times 10^{28}\ \mathrm{GeV}^{-1}$.

Define 
\begin{equation}
    n_0(E) \, = \, \frac{1}{V}\int_E^\infty N_0(E')\mathrm{d}E'
    \label{eqn:densdef}
\end{equation}
\noindent where $V$ is the total volume of the trapping region, with $n_0(E)$  the average density of accelerated ions of energy $E$ or greater. 
Scenarios in which $n_0$ is not implausibly large are preferred.

The flare seen in 1600~\AA\ and 1700~\AA\ AIA channels has a characteristic scale of $\sim 45''$ once its morphology stabilizes. Supposing, e.g. that the fast ion reservoir is a hemisphere of radius $22''$ leads to a volume of $8.5\times 10^{27}\ \mathrm{cm}^3$. 
The density of fast ions is highly sensitive to the value of the lowest energy for which the power-law energy distribution still holds, a quantity that is only poorly constrained by observation. 
Studies of deexcitation lines often use a value of 30~MeV \citep[e.g.][]{1984AdSpR...4g.127M} while 1~MeV has been used for estimates of the flare energy budget \citep{Ems:al-05}. 
The former case, which is compatible with the limit to the flux in the GBM energy range (Sect.~\ref{sec:gbm_analysis}), leads to $n_0(.03) = 4500\ \mathrm{cm}^{-3}$, a plausible value, but  $n_0(0.001) = 4.7 \times10^8 \ \mathrm{cm}^{-3}$. 
The latter value is probably too close to the total ambient density in the low corona to be acceptable. 
We conclude that this gradual escape model is plausible as long as the energetic ion distribution flattens below 30~MeV, or the relevant volume is greater than assumed here.

\section{Conclusions}
\label{sec:conclusions}
In this work we have presented multiwavelength observations of SOL2012-06-03 with particular attention to the $>$100~MeV $\gamma$-ray peaks during the impulsive phase.  As outlined in Section~\ref{sec:interpretation}, the data strongly suggest that there are multiple acceleration sites and that the first impulsive peak is a flare-related event coincident with the radio and HXR emissions. Whereas the particle acceleration during the predominately ion-rich second $\gamma$-ray peak can be due to an impulsive burst of reconnection, or a flux-rope driven shock that moves through the weak-field region (although it is uncertain whether this process can develop over very short times scales).  The delayed emission observed in $\gamma$-rays following the impulsive phase can be explained by trapping of ions in closed field lines and precipitation. This flare represents the first time that two distinct phases are observed for $>$100~MeV emission and that these phases clearly indicate separate acceleration process and locations of the energy releases of the accelerated particles.

\begin{acknowledgments}
The \textit{Fermi} LAT Collaboration acknowledges generous ongoing support
from a number of agencies and institutes that have supported both the
development and the operation of the LAT as well as scientific data analysis.
These include the National Aeronautics and Space Administration and the
Department of Energy in the United States, the Commissariat \`a l'Energie Atomique
and the Centre National de la Recherche Scientifique / Institut National de Physique
Nucl\'eaire et de Physique des Particules in France, the Agenzia Spaziale Italiana
and the Istituto Nazionale di Fisica Nucleare in Italy, the Ministry of Education,
Culture, Sports, Science and Technology (MEXT), High Energy Accelerator Research
Organization (KEK) and Japan Aerospace Exploration Agency (JAXA) in Japan, and
the K.~A.~Wallenberg Foundation, the Swedish Research Council and the
Swedish National Space Board in Sweden.
Additional support for science analysis during the operations phase is gratefully
acknowledged from the Istituto Nazionale di Astrofisica in Italy and the Centre
National d'\'Etudes Spatiales in France.  MPR and NO are members of the LAT collaboration. SDO data were supplied courtesy of the SDO/HMI and SDO/AIA consortia. SDO is the first mission to be launched for NASA's Living With a Star (LWS) Program. This work was performed in part under DOE
Contract DE-AC02-76SF00515. 

We are grateful for radio data of the Bleien (Switzerland) and Birr Castle (Ireland) instruments of the e-CALLISTO network and of the POEMAS radio polarimeters of the Centro de R\'adio-Astronomia e Astrofísica Mackenzie (Brazil) operated at the CASLEO observatory in Argentina (data provided by G. Gim\'enez de Castro).

TW acknowledges DLR grant 50 OC 2301. AJBR and CEP acknowledge support from STFC grants ST/S000402/1 and ST/W001195/1. 
HSH is grateful for hospitality at the University of Glasgow.
This research used version 6.0.2 \citep{sunpy_community2020} of the SunPy open source software package \citep{sunpy-project}.
 
NVN acknowledges NASA grants 80NSSC23K0408 and 80NSSC24K0175.

\end{acknowledgments}

\appendix
\section{\emph{Fermi}-LAT time resolved spectral analysis}
\label{app:tables}
In this section we report the \emph{Fermi}-LAT time resolved spectral analysis results in Table~\ref{tab:spectral_results}.

\begin{table}[h!]
    \centering
    \begin{tabular}{c|c c  c}
    Time window & Flux$_{0.1-1.0\rm GeV}$ & power-law index & TS$_{sun}$ \\
              &  (10$^{-5}$cm$^{-2}$ s$^{-1}$)   &  &   \\
    \hline
    \hline
    17:53:10 -- 17:53:18 & $ 16 \pm 5 $  & $ -2.4 \pm 0.4 $ &  80\\
17:53:18 -- 17:53:21 & $ 43 \pm 14 $  & $ -2.6 \pm 0.4 $  & 84\\
17:53:21 -- 17:53:24 & $ 38 \pm 12 $  & $ -2.9 \pm 0.4 $  & 122\\
17:53:24 -- 17:53:28 & $ 29 \pm 9 $  & $ -2.4 \pm 0.4 $  & 85\\
17:53:28 -- 17:53:32 & $ 30 \pm 11 $  & $ -2.5 \pm 0.4 $  & 80\\
17:53:32 -- 17:53:34 & $ 98 \pm 27 $  & $ -2.20 \pm 0.30 $ &  121\\
17:53:34 -- 17:53:36 & $ 60 \pm 19 $  & $ -2.5 \pm 0.4 $  & 83\\
17:53:36 -- 17:53:37 & $ 104 \pm 31 $  & $ -2.4 \pm 0.4 $ &  130\\
17:53:37 -- 17:53:40 & $ 48 \pm 16 $  & $ -2.9 \pm 0.5 $  & 88\\
17:53:40 -- 17:53:41 & $ 83 \pm 24 $  & $ -2.4 \pm 0.4 $  & 132\\
17:53:41 -- 17:53:46 & $ 26 \pm 7 $  & $ -2.13 \pm 0.31 $  & 94\\
17:53:46 -- 17:54:01 & $ 7.6 \pm 2.5 $  & $ -2.8 \pm 0.5 $  & 70\\
17:54:01 -- 17:54:18 & $ 5.6 \pm 2.1 $  & $ -1.86 \pm 0.33 $ &  51\\
17:54:18 -- 17:54:40 & $ 5.2 \pm 1.8 $  & $ -2.3 \pm 0.4 $  & 36\\
17:54:40 -- 17:55:04 & $ 4.7 \pm 1.6 $  & $ -2.01 \pm 0.32 $ &  48\\
17:55:04 -- 17:55:34 & $ 3.3 \pm 1.2 $  & $ -2.6 \pm 0.5 $  & 37\\
17:55:34 -- 17:55:53 & $ 5.0 \pm 1.8 $  & $ -2.5 \pm 0.4 $  & 42\\
17:55:53 -- 17:56:28 & $ 3.2 \pm 1.1 $  & $ -1.85 \pm 0.28 $ &  45\\
17:56:28 -- 17:56:55 & $ 3.0 \pm 1.3 $  & $ -3.1 \pm 0.6 $  & 28\\
17:56:55 -- 17:57:35 & $ 2.2 \pm 0.8 $  & $ -2.7 \pm 0.5 $  & 27\\
17:57:35 -- 17:58:09 & $ 2.7 \pm 1.0 $  & $ -2.8 \pm 0.5 $  & 30\\
17:58:09 -- 17:58:39 & $ 3.4 \pm 1.2 $  & $ -2.6 \pm 0.4 $  & 38\\
17:58:39 -- 17:58:58 & $ 5.1 \pm 1.9 $  & $ -2.3 \pm 0.4 $  & 31\\
17:58:58 -- 17:59:36 & $ 2.3 \pm 0.9 $  & $ -2.02 \pm 0.35 $ &  23\\
17:59:36 -- 18:00:18 & $ 2.4 \pm 0.9 $  & $ -2.5 \pm 0.4 $  & 33\\
18:00:18 -- 18:01:21 & $ 1.2 \pm 0.5 $  & $ -2.1 \pm 0.4 $  & 17\\
18:01:21 -- 18:02:11 & $ 1.6 \pm 0.7 $  & $ -2.2 \pm 0.4 $  & 19\\
\hline
    \end{tabular}
      \caption{Results for the time resolved likelihood analysis of the LAT data. In each interval, we report the time window for the analysis, the photon flux between 100~MeV and 1~GeV (in 10$^{-5}$cm$^{-2}$ s$^{-1}$), the photon power-law index from the fit and the Test Statistic, TS, from the fit (significance $= \sqrt{TS}$). }
    \label{tab:spectral_results}
\end{table}

\newpage

\bibliography{SOL20120603}
\bibliographystyle{aasjournal}

\end{document}